\documentclass[useAMS,usenatbib]{mn2e}
\usepackage{graphicx}
\usepackage{color}
\usepackage{soul}
\usepackage{bm}
\usepackage{amssymb,amsfonts,amsmath}%

\newcommand{\bec}[1]{\mbox{\boldmath $ #1$}}
\newcommand{\meanrho}{\overline{\rho}}
\newcommand{\meanAA}{\overline{\mbox{\boldmath $A$}}{}}{}
\newcommand{\meanBB}{\overline{\mbox{\boldmath $B$}}{}}{}
\newcommand{\meanUU}{\overline{\bm{U}}}

{}
\newcommand{\meanA}{\overline{A}}
\newcommand{\meanB}{\overline{B}}

\newcommand{\meanU}{\overline{U}}

\newcommand{\meanP}{\overline{P}}

\newcommand{\meanT}{\overline{T}}

\title[Budget equations and astrophysical nonlinear mean-field dynamos]
{Budget equations and astrophysical nonlinear mean-field dynamos}

\author[
I. Rogachevskii, N. Kleeorin
]
{I. Rogachevskii,$^{1,2}$
N. Kleeorin,$^{1,3}$
\\
 $^{1}$ Department of Mechanical Engineering,
        Ben-Gurion University of Negev, POB 653, Beer-Sheva  84105, Israel\\
 $^{2}$ Nordita, KTH Royal Institute of Technology and Stockholm University,
        Roslagstullsbacken 23, Stockholm  SE-10691, Sweden\\
 $^{3}$ Institute of Continuous Media Mechanics, Korolyov str. 1, Perm  614013, Russia}

\begin{document}


\maketitle


\begin{abstract}
Solar, stellar and galactic large-scale magnetic fields are originated due to a combined action of non-uniform (differential) rotation and helical motions of plasma via mean-field dynamos. Usually, nonlinear mean-field dynamo theories take into account algebraic and dynamic quenching of alpha effect and algebraic quenching of turbulent magnetic diffusivity.
However, the theories of the algebraic quenching do not take into account
the effect of modification of the source of turbulence by the growing large-scale  magnetic field.
This phenomenon is due to the dissipation of the strong large-scale magnetic field
resulting in an increase of the total turbulent energy.
This effect has been studied using the budget equation for the total turbulent energy
(which takes into account the feedback of the generated large-scale magnetic field
on the background turbulence)
for (i) a forced turbulence, (ii) a shear-produced turbulence and (iii) a convective turbulence.
As the result of this effect, a nonlinear dynamo number decreases with increase
of the large-scale magnetic field, so that that the mean-field $\alpha\Omega$, $\alpha^2$ and $\alpha^2\Omega$
dynamo instabilities are always saturated by the strong large-scale magnetic field.
\end{abstract}


\maketitle


\section{Introduction}
\label{sect-1}

Large-scale magnetic fields in the Sun, stars and galaxies are believed to be generated by
a joint action of a differential rotation and helical motions of plasma
\citep[see, e.g.,][]{Moffatt(1978),Parker(1979),Krause(1980),Zeldovich(1983),Ruzmaikin(1988),Ruediger(2013),MD2019,RI21,SS21}.
This mechanism can be described by the $\alpha\Omega$ or $\alpha^2\Omega$ mean-field dynamos.
In particular, the effect of turbulence in the mean-field induction equation
is determined by the turbulent electromotive force, $\langle {\bm u} \times {\bm b} \rangle$,
which can be written for a weak mean magnetic field $\meanBB$
as $\langle {\bm u} \times {\bm b} \rangle= \alpha_{_{\rm K}} \,  \meanBB
+ {\bm V}^{({\rm eff})}\times \meanBB - \eta_{_{T}} \, ({\bm \nabla} \times \meanBB)$,
where $\alpha_{_{\rm K}}$ is the kinetic $\alpha$ effect
caused by helical motions of plasma,
$\eta_{_{T}}$ is the turbulent magnetic diffusion coefficient,
${\bm V}^{({\rm eff})}$ is the effective pumping velocity caused by an inhomogeneity of turbulence.
Here the angular brackets imply ensemble averaging, ${\bm u}$ and ${\bm b}$ are
fluctuations of velocity and magnetic fields, respectively.
The threshold of the $\alpha\Omega$ mean-field dynamo instability
is described in terms of a dynamo number
$D_{\rm L}=\alpha_{_{\rm K}} \, \delta \Omega \, L^3/ \eta_{_{T}}^2$,
where  $\delta \Omega$ characterises the non-uniform (differential) rotation
and $L$ is the stellar radius or the thickness of the galactic disk.

The mean-field dynamos are saturated by nonlinear effects.
In particular, a feedback of the growing large-scale magnetic field on plasma motions is
described by algebraic quenching of the $\alpha$ effect, turbulent magnetic diffusion,
and the effective pumping velocity.
This implies that the turbulent transport coefficients, $\alpha_{_{\rm K}}\left(\meanB\right)$,
$\eta_{_{T}}\left(\meanB\right)$ and ${\bm V}^{({\rm eff})}\left(\meanB\right)$
depend on the mean magnetic field $\meanBB$ via algebraic decreasing functions.
The quantitative theories of the algebraic nonlinearities of the
$\alpha$ effect, the turbulent magnetic diffusion and the effective pumping velocity
have been developed using the quasi-linear approach
for small fluid and magnetic Reynolds numbers
\citep{RK93,KRP94,Ruediger(2013)}
and the tau approach for large fluid and magnetic Reynolds numbers
\citep{FBC99,RK2000,RK2001,RK2004,RK2006}.

In addition to the algebraic nonlinearity, there is also a dynamic nonlinearity
caused by an evolution of magnetic helicity density of a small-scale turbulent magnetic field
during the nonlinear stage of the mean-field dynamo.
Indeed, the $\alpha$ effect has contributions from the kinetic $\alpha$ effect, $\alpha_{_{\rm K}}$,
determined by the kinetic helicity and a magnetic $\alpha$ effect, $\alpha_{_{\rm M}}$,
described by the current helicity of the small-scale magnetic field \citep{PFL76}.
The dynamics of the current helicity are determined by the
evolution of the small-scale magnetic helicity density
$H_{\rm m} =\langle {\bm a} {\bf \cdot} {\bm b} \rangle$, where
${\bm b}=\bec{\nabla} {\bf \times} {\bm a}$ and ${\bm a}$ are fluctuations of the magnetic vector potential.
The total magnetic helicity, i.e., the sum of the magnetic helicity densities
of the large-scale and small-scale magnetic fields, $H_{\rm M}+H_{\rm m}$,
integrated over the volume, $\int (H_{\rm M} + H_{\rm m}) \,dr^3$,
is conserved for very small microscopic magnetic diffusivity $\eta$.
Here $H_{\rm M}=\overline{\bm A} {\bf \cdot} \overline{\bm B}$
is the magnetic helicity density of the large-scale magnetic field $\overline{\bm B}
=\bec{\nabla} {\bf \times} \overline{\bm A}$ and $\overline{\bm A}$ is the mean magnetic vector potential.

As the mean-field dynamo instability amplifies the mean magnetic field,
the large-scale magnetic helicity density $H_{\rm M}$ grows in time.
Since the total magnetic helicity $\int (H_{\rm M} + H_{\rm m}) \,dr^3$
is conserved for very small magnetic diffusivity, the magnetic helicity
density $H_{\rm m}$ of the small-scale field changes during the dynamo action,
and its evolution is determined by the dynamic equation
\citep{KR82,Zeldovich(1983),GD94,KRR95,KR99},
which includes the source terms
and turbulent fluxes of magnetic helicity
\citep{KR99,KMR00,BF00,VC01,Brandenburg(2005),KR22,GS23}.

Taking into account turbulent fluxes of the small-scale magnetic helicity,
it has been shown by numerical simulations that a nonlinear galactic dynamo
governed by a dynamic equation for the
magnetic helicity density $H_{\rm m}$ of a small-scale field (the dynamical nonlinearity)
saturates at a mean magnetic field comparable with the
equipartition magnetic field
\citep[see, e.g.,][]{KMR00,KMR02,KMR03a,BB02,Brandenburg(2005),SSS06,CSS14,CS18}.
Numerical simulations demonstrate that the dynamics of magnetic helicity
plays a crucial role in solar and stellar dynamos as well
\citep[see, e.g.,][]{KKMR03,KSR16,KSR20,KRS23,SKR06,ZKRS06,ZKRS12,KKB10,HB12,DGB13,SKR18,RIN21}.
Different forms of magnetic helicity fluxes have been suggested in various studies using phenomenological arguments
\citep{KR99,KMR00,KMR02,VC01,SB04,Brandenburg(2005)}.
Recently, the turbulent magnetic helicity fluxes have been rigorously derived  \citep{KR22,GS23}.
In particular, \cite{KR22} apply the mean-field theory, adopt the Coulomb gauge and
consider a strongly density-stratified turbulence.
They have found that the turbulent magnetic helicity fluxes depend on the
mean magnetic field energy,  and include
non-gradient and gradient contributions.
In addition, \cite{GS23} have recently shown that
contributions to the turbulent magnetic helicity fluxes from the third-order moments
can be described using the turbulent diffusion approximation.

In a nonlinear $\alpha\Omega$ dynamo one can define
a nonlinear dynamo number
$D_{\rm N}\left(\meanB\right) = \alpha\left(\meanB\right) \,
\delta \Omega \, L^3/ \eta_{_{T}}^2\left(\meanB\right)$.
If the nonlinear dynamo number $D_{\rm N}\left(\meanB\right)$
decreases with the increase of the large-scale magnetic field,
the mean-field dynamo instability is saturated by the nonlinear effects.
However, if the $\alpha$ effect and the turbulent magnetic diffusion are quenched as
$(\meanB/ \meanB_{\rm eq})^{-2}$ for strong mean magnetic fields,
the nonlinear dynamo number $D_{\rm N}\left(\meanB\right) \propto (\meanB/ \meanB_{\rm eq})^{2}$
increases with the increase of the large-scale magnetic field, and the mean-field dynamo instability
cannot be saturated for a strong mean magnetic field.
Here $\meanB_{\rm eq} = \left(\mu_0\, \meanrho \, \langle
{\bm u}^2 \rangle\right)^{1/2}$ is the equipartition mean magnetic field
and $\mu_0$ is the magnetic permeability of the fluid.
How is it possible to resolve this paradox?

The mean-field dynamo theories of the algebraic quenching
imply that there is a background helical turbulence
with a zero mean magnetic field.
The large-scale magnetic field
is amplified by the mean-field dynamo instability.
In a nonlinear dynamo stage, the dissipation of the generated strong
large-scale magnetic field results in an increase of the turbulent kinetic energy
of the background turbulence.
The latter effect causes an increase of the turbulent magnetic diffusion
coefficient and decrease of the nonlinear dynamo number.
This additional nonlinear effect results in a saturation of the dynamo growth of a strong
large-scale magnetic field.

However, this nonlinear effect has not been yet taken into account
in nonlinear mean-field dynamo theories
which derive the algebraic quenching of the turbulent
magnetic diffusion.
In the present study, we have taken into account this feedback effect
of the mean magnetic field on the background turbulence
using the budget equation for the total (kinetic plus magnetic) turbulent energy.
Considering three different types of astrophysical turbulence:
\begin{itemize}
\item{
a forced turbulence (e.g., caused by supernova explosions in galaxies);}
\item{
a shear-produced turbulence (e.g., in the atmosphere of the Earth or other planets)
and}
\item{
a convective turbulence (e.g., in a solar and stellar convective zones),}
\end{itemize}
we have demonstrated that the nonlinear dynamo number indeed decreases
with the increase of the mean magnetic field
for any strong values of the magnetic field,
resulting in saturation of the mean-field dynamo instability.

This paper is organized as follows.
In Sec.~\ref{sect-2} we
explain the essence of the algebraic and dynamic nonlinearities,
and discuss the procedure and assumptions for the derivation
of the nonlinear turbulent electromotive force (EMF).
In Sec. ~\ref{sect-3} we consider the budget equations for the turbulent kinetic and magnetic energies
which allow us to take into account the increase of turbulent kinetic energy of the background turbulence by
the dissipation of a strong mean magnetic field
and to determine asymptotic properties of turbulent magnetic diffusion
and nonlinear dynamo numbers for a strong mean magnetic field
for the mean-field $\alpha \, \Omega$ dynamo
(see Sec. ~\ref{sect-4}), the $\alpha^2$ dynamo
(see Sec. ~\ref{sect-5}) and the $\alpha^2 \, \Omega$ dynamo
(see Sec. ~\ref{sect-6}).
In addition, in Sec. ~\ref{sect-5} we discuss a long-standing question when a kinematic
$\alpha^2$ dynamo can be oscillatory, and in Sec. ~\ref{sect-6} we outline
important asymptotic properties in the $\alpha^2 \, \Omega$ dynamo.
Finally,  in Sec.~\ref{sect-7} we discuss the obtained results.

\section{Nonlinear turbulent EMF}
\label{sect-2}

To explain the essence of the algebraic and dynamic nonlinearities,
we discuss in this section the procedure and assumptions for the derivation
of the nonlinear turbulent electromotive force in
a non-rotating and helical small-scale turbulence.
In the framework of the mean-field approach, the mean magnetic
field $\meanBB$ is determined by the induction equation
\begin{eqnarray}
{\partial \meanBB \over \partial t} = \bec{\nabla} \times
\left[\meanUU {\bm \times} \meanBB + \bec{\cal E}\left(\meanBB\right)
- \eta \bec{\nabla} {\bm \times} \meanBB\right] ,
\label{WEI1}
\end{eqnarray}
where $\meanUU$ is the mean velocity (differential rotation), $\eta$ is the magnetic
diffusion due to the electrical conductivity of plasma and
$\bec{\cal E}(\meanBB) = \langle {\bm u}
\times {\bm b} \rangle$ is the the turbulent
electromotive force (EMF).
To derive equations for the nonlinear
coefficients defining the turbulent electromotive force (EMF), we use a
mean-field approach in which the magnetic and velocity fields,
the fluid pressure and density are
separated into the mean and fluctuating parts, where the fluctuating
parts have zero mean values. We consider the case of large
hydrodynamic and magnetic Reynolds numbers. The momentum
and induction equations for the turbulent fields are given by
\begin{eqnarray}
&& {\partial {\bm u}(t,{\bm x}) \over \partial t} = - {\bec{\nabla}
p_{\rm tot} \over \meanrho} + {1 \over \mu_0 \meanrho} \left[({\bm b} \cdot
\bec{\nabla}) \meanBB +\left (\meanBB \cdot \bec{\nabla}\right){\bm b}\right]
\nonumber\\
&&\qquad+ {\bm u}^N + {\bm F} ,
\label{WEB1} \\
&& {\partial {\bm b}(t,{\bm x}) \over \partial t} = \left(\meanBB
\cdot \bec{\nabla}\right){\bm u} - ({\bm u} \cdot \bec{\nabla}) \meanBB + {\bm b}^N ,
\label{WEB2}
\end{eqnarray}
where $\meanrho$ is the mean plasma density,
$\mu_0$ is the magnetic permeability of the plasma,
${\bm F}$ is a random external stirring force, ${\bm u}^{N}$ and
${\bm b}^{N}$ are the nonlinear terms which include the molecular
dissipative terms, $p_{\rm tot} = p + (\mu_0\, \meanrho\, )^{-1} \,(\meanBB
\cdot {\bm b})$ are fluctuations of the total pressure and $p$ are
fluctuations of the plasma pressure. For simplicity, let us consider incompressible flow,
so that the velocity ${\bm u}$ satisfies to the continuity equation,
$\bec{\nabla} \cdot {\bm u} = 0$ and the fluid density is constant.
The assumptions and the procedure of the derivation of
the nonlinear turbulent electromotive force are as follows.
\begin{itemize}
\item{
We apply the multi-scale approach
\citep{RS75}, which allows us to introduce fast and slow variables, and separate small-scale
effects corresponding to fluctuations and large-scale effects describing mean fields.
The mean fields depend on slow variables, while fluctuations
depend on fast variables.
Separation into slow and fast variables is widely used in theoretical physics,
and all calculations are reduced to the Taylor expansions of all functions
assuming that characteristic turbulent spatial and time scales are much smaller than
the characteristic spatial and time scales of the mean magnetic field variations.
}
\item{
Using Eqs.~(\ref{WEB1})--(\ref{WEB2}) written in a Fourier space, we
derive equations for the second-order moments for the velocity
field $f_{ij}=\langle u_i u_j \rangle$, the magnetic field
$h_{ij}=\langle b_i b_j \rangle$ and the cross-helicity
$g_{ij}=\langle u_i b_j \rangle$.}
\item{
We split the tensors $f_{ij}$, $h_{ij}$ and $g_{ij}$
into nonhelical $h_{ij}$ and helical, $h_{ij}^{(H)}$ parts. The
helical part of the tensor $h_{ij}^{(H)}$ for magnetic
fluctuations depends on the small-scale magnetic helicity, and its evolution is
determined by the dynamic equation which follows from the magnetic
helicity conservation arguments
\citep{KR82,GD94,KRR95,KR99,KMR00,BB02}. The characteristic
time of the evolution of the nonhelical part of the magnetic tensor
$h_{ij}$ is of the order of the turbulent correlation time
$\tau_{0} = \ell_{0} / u_{0}$, while the relaxation time of the
helical part of the magnetic tensor $h_{ij}^{(H)}$ is of the
order of $\tau_{0} \,{\rm Rm}$,
where ${\rm Rm} = \ell_0 u_{0} / \eta \gg 1 $ is the magnetic
Reynolds number, and $u_{0}$ is the characteristic turbulent velocity
in the integral scale $\ell_{0}$ of turbulent motions.}
\item{
Equations for the second-order moments contain higher-order moments and a
problem of closing the equations for the higher-order moments arises.
Various approximate methods have been proposed for the solution of
this closure problem \citep{MY71,MY13,MC90,RI21}.
For small fluid and magnetic Reynolds numbers,
the quasi-linear approach can be used
\citep{RK93,KRP94,Ruediger(2013)},
while for large fluid and magnetic Reynolds numbers, the minimal tau approach \citep{FBC99}
or the spectral $\tau$ approach \citep{RK2000,RK2001,RK2004,RK2006} are applied
to derive the nonlinear turbulent electromotive force.
For instance, the spectral $\tau$ approach postulates that the deviations of the third-order moments,
$\hat{\cal M} f_{ij}^{(III)}({\bm k})$, from the contributions to these terms afforded by the background turbulence, $\hat{\cal M} f_{ij}^{(III,0)}({\bm k})$,
can be expressed through the similar deviations of the second-order moments,
$f_{ij}^{(II)}({\bm k}) - f_{ij}^{(II,0)}({\bm k})$ \citep{O70,PFL76,KRR90}:
\begin{eqnarray}
\hat{\cal M} f_{ij}^{(III)}({\bm k}) - \hat{\cal M} f_{ij}^{(III,0)}({\bm
k}) = - {f_{ij}^{(II)}({\bm k}) - f_{ij}^{(II,0)}({\bm k}) \over \tau_r(k)}  ,
\label{WEA1}
\end{eqnarray}
where $\tau_r(k)$ is the scale-dependent relaxation time, which can be
identified with the correlation time $\tilde \tau(k)$ of the turbulent velocity field
for large fluid and magnetic Reynolds numbers.
The superscript ${(0)}$ corresponds to the
background turbulence (with $ \meanBB = 0)$, and
$\tau_r(k)$ is the characteristic relaxation time of the statistical moments.
We apply the spectral $\tau$ approach only for the nonhelical part
$h_{ij}$ of the tensor for magnetic fluctuations.
The spectral $\tau$ approach is widely
used in the theory of kinetic equations, in passive scalar
turbulence and magnetohydrodynamic turbulence.}
\item{
We use the following model for the second-order moment $f_{ij}^{(0)}$
of isotropic inhomogeneous incompressible and helical background
turbulence in a Fourier space:
\begin{eqnarray}
&& f_{ij}^{(0)}({\bm k}) = {E(k) \over 8 \pi k^2} \, \Big\{ \Big[\delta_{ij} - k_{ij} + {{\rm i} \over 2 k^2} \, \big(k_i \nabla_j
\nonumber\\
&& \quad - k_j \nabla_i\big) \Big] \left\langle {\bm u}^2 \right\rangle^{(0)}
- {{\rm i} \over k^2} \, \varepsilon_{ijp} \, k_p \, \left\langle{\bm u} \,{\bm \cdot}  \,(\bec{\nabla} {\bm \times} \, {\bm u}) \right\rangle \Big\} .
\label{WEM13}
\end{eqnarray}
Here $\delta_{ij}$ is the Kronecker tensor, $k_{ij} = k_i \, k_j /k^2$ and
$\langle{\bm u} \,{\bm \cdot}  \,(\bec{\nabla} {\bm \times} \, {\bm u}) \rangle$
is the kinetic helicity.
The energy spectrum function is $E(k) = (2/3)  \, k_0^{-1} \, (k / k_{0})^{-5/3}$ in the inertial range of turbulence $k_0 \leq k \leq k_{\nu}$. Here the wave number $k_{0} = 1 / \ell_0$, the length $\ell_0$ is the integral scale of turbulent motions, the wave number $k_{\nu}=\ell_{\nu}^{-1}$, the length $\ell_{\nu} = \ell_0 {\rm Re}^{-3/4}$ is the Kolmogorov (viscous) scale, and the expression for the turbulent correlation time is given by $\tilde \tau(k) = 2 \, \tau_0 \, (k / k_{0})^{-2/3}$.
The model for the second moment $h_{ij}^{(0)}$
for magnetic fluctuations in a Fourier space caused by the small-scale dynamo
(with a zero mean magnetic field) is
\begin{eqnarray}
h_{ij}^{(0)}({\bm k}) &=& {E(k) \over 8 \pi k^2} \Big(\delta_{ij} - k_{ij}\Big) \left\langle {\bm b}^2 \right\rangle^{(0)} .
\label{WEMM13}
\end{eqnarray}
We also take into account that the turbulent electromotive force is
produced in a turbulence with a non-zero mean magnetic field, so that the cross-helicity tensor in the background turbulence vanishes, i.e., $g_{ij}^{(0)}=0$.
}
\item{
We assume that the characteristic time of variation of the mean magnetic
field $\meanBB$ is substantially larger than the correlation
time $\tilde \tau(k)$ for all turbulence scales (which corresponds to the
mean-field approach). This allows us to get a stationary solution
for the equations for the second moments.
Using the derived equations for the second moments $f_{ij}$,
$h_{ij}$ and $g_{ij}$, we determine the nonlinear turbulent electromotive force
${\cal E}_{i} = \varepsilon_{imn} \int g_{mn}({\bm k}) \,d {\bm k}$.
The details of the derivation of the nonlinear turbulent electromotive force are given by \cite{RK2004}.}
\end{itemize}

For illustration of these results, we consider a small-scale homogeneous turbulence with a mean
velocity shear, $ \meanUU = S \, z \, {\bm e}_y $.
We also consider, an axi-symmetric $\alpha\Omega$ dynamo
problem in the cartesian coordinates, so  the mean
magnetic field, $\meanBB = \meanB_{y}(x,z) \, {\bm e}_y + \bec{\nabla}
{\bm \times} [\meanA(x,z) \, {\bm e}_y]$, is determined by the following
nonlinear dynamo equations \citep{RK2004}:
\begin{eqnarray}
{\partial \meanA \over \partial t} &=&
\left[\alpha_{_{\rm K}}(\meanB) + \alpha_{_{\rm M}}(\meanBB) \right] \, \, \meanB_{y}
+\eta_{_{T}}^{(A)}\left(\meanBB\right) \, \Delta \meanA ,
\label{WEF11} \\
{\partial \meanB_{y} \over \partial t} &=& S \, \nabla_x  \, \meanA + \nabla_j \left[\eta_{_{T}}^{(B)}\left(\meanBB\right)
\nabla_j \right]  \meanB_{y} .
\label{WEF12}
\end{eqnarray}
Here, the nonlinear $\alpha$ effect is given by
\begin{eqnarray}
\alpha(\meanBB) &=& \alpha_{_{\rm K}}(\meanB) + \alpha_{_{\rm M}}(\meanBB) ,
\label{WEL20}
\end{eqnarray}
where $\alpha^{({\rm K})}(\meanB)$ is the kinetic $\alpha$ effect,
and $\alpha^{({\rm M})}\left(\meanBB\right) $ is the magnetic $\alpha$ effect, which are given by
\begin{eqnarray}
\alpha_{_{\rm K}}\left(\meanB\right) &=& \alpha_{_{\rm K}}^{(0)} \, \phi_{_{\rm K}}(\beta) \, (1 - \epsilon) ,
\label{WEPL1}\\
\alpha_{_{\rm M}}\left(\meanB\right)  &=& {\tau_0 \over 3 \mu_0\, \overline{\rho}} \, H_{\rm c}\left(\meanBB\right) \,  \phi_{_{\rm M}}(\beta)  .
\label{WEPL2}
\end{eqnarray}
Here $\alpha_{_{\rm K}}^{(0)} = - \tau_0 H_{\rm u}/3$ with
$H_{\rm u} = \langle {\bm u} {\bf \cdot} (\bec{\nabla} {\bf \times} {\bm u}) \rangle$ being the kinetic helicity,
$\beta= \sqrt{8} \, \meanB/\meanB_{\rm eq}$,
the parameter $\epsilon=\langle {\bm b}^2 \rangle^{(0)} \, \ell_b/(\langle {\bm u}^2 \rangle^{(0)} \ell_0)$
characterised the small-scale dynamo is varied in the range $0 \leq \epsilon \leq 1$,
where $\langle {\bm b}^2 \rangle^{(0)}/2\mu_0$ and $\langle {\bm u}^2 \rangle^{(0)}/2$
are turbulent magnetic and kinetic energies of the background turbulence, $\ell_b$
is the characteristic scale of the localization of the magnetic energy due to the small-scale dynamo,
and $H_{\rm c}\left(\meanBB\right)=\langle {\bm b} {\bf \cdot} (\bec{\nabla} {\bf \times} {\bm b}) \rangle$
is the current helicity of the small-scale magnetic field ${\bm b}$.

The quenching functions $\phi_{_{\rm K}}(\beta)$ and $\phi_{_{\rm M}}(\beta)$ of the kinetic and magnetic $\alpha$ effects are
given by Eqs.~(\ref{WECB41})--(\ref{WECB40})  in Appendix~\ref{sect-A1}.
Here $\phi_{_{\rm M}}(\beta)$ is the quenching function of the
magnetic $\alpha$ effect derived by \cite{FBC99} using the minimal $\tau$ approximation
(the $\tau$ approach applied in a physical space)
and  \cite{RK2000} using the spectral $\tau$ approach.

The nonlinear turbulent magnetic diffusion coefficients for the poloidal $\eta_{_{T}}^{(A)}\left(\meanB\right)$ and toroidal $\eta_{_{T}}^{(B)}\left(\meanB\right)$ mean magnetic field are given by
\begin{eqnarray}
\eta_{_{T}}^{(A)}(\beta)  &=&  \eta_{_{T}}^{(0)} \, \phi_{_{\rm K}}(\beta)  , \quad  \eta_{_{T}}^{(B)}(\beta)  =  \eta_{_{T}}^{(0)} \, \phi_\eta^{(B)}(\beta) ,
\label{WEB41}
\end{eqnarray}
where $\eta_{_{T}} = \tau_0 \, \left\langle {\bm u}^2 \right\rangle /3$ is the characteristic
value of the turbulent magnetic diffusivity.
The quenching function $\phi_\eta^{(B)}(\beta)=\phi_{_{\rm K}}(\beta) + \phi(\beta)$ and
the functions $\phi_{_{\rm K}}(\beta)$ and $\phi(\beta)$ are given by
Eqs.~(\ref{WECB41}) and~(\ref{WECB402}) in Appendix~\ref{sect-A1}.
Here for simplicity we consider a homogeneous background turbulence, so
the effective pumping velocity of the large-scale magnetic field vanishes.

The asymptotic formulas for the kinetic and magnetic
$\alpha$ effects, and the nonlinear turbulent magnetic diffusion
coefficients of the mean
magnetic field for a weak field $\meanB \ll \meanB_{\rm eq} / 4$ are given by
\begin{eqnarray}
\alpha^{({\rm K})}(\beta) &=& \alpha_{_{\rm K}}^{(0)} \,(1 - \epsilon) \,
\biggr(1 - {12 \beta^{2} \over 5}\biggr) ,
\label{PB1}\\
\alpha^{({\rm M})}\left(\meanBB\right) &=& {\tau_0 \over 3 \mu_0\, \overline{\rho}} \, H_{\rm c}\left(\meanBB\right) \,
\biggr(1 - {3 \beta^{2} \over 5}\biggr) ,
\label{PB2}\\
\eta_{_{T}}^{(A)}(\beta) &=&  \eta_{_{T}}^{(0)} \, \left(1 - {12 \over 5} \, \beta^{2}\right) ,
\label{PB3}\\
\eta_{_{T}}^{(B)}(\beta) &=& \eta_{_{T}}^{(0)} \, \left(1 - {4 \over 5} \, (5 - 4\epsilon) \,
\beta^{2}\right) ,
\label{PB4}
\end{eqnarray}
and for a strong field $\meanB \gg \meanB_{\rm eq} / 4$ they are given by
\begin{eqnarray}
\alpha^{({\rm K})}(\beta) &=&  {\alpha_{_{\rm K}}^{(0)} \over \beta^{2}} \,(1 - \epsilon) ,
\label{PB5}\\
\alpha^{({\rm M})}\left(\meanBB\right) &=& {\tau_0 \over  \mu_0\, \overline{\rho}} \,
\, {H_{\rm c}\left(\meanBB\right) \over \beta^{2}} ,
\label{PB6}\\
\eta_{_{T}}^{(A)}(\beta) &=&  {\eta_{_{T}}^{(0)}  \over \beta^2} , \quad
\eta_{_{T}}^{(B)}(\beta) =  {2 \eta_{_{T}}^{(0)} \over 3 \beta}  \, (1 + \epsilon) .
\label{PB7}
\end{eqnarray}
It follows from Eqs.~(\ref{PB1})--(\ref{PB7}), that small-scale dynamo decreases the kinetic $\alpha$ effect and
it increases the turbulent magnetic diffusion of the toroidal mean magnetic field.

As follows from Eq.~(\ref{WEPL2}),
the magnetic $\alpha$ effect is proportional to the current helicity
$H_{\rm c}\left(\meanBB\right)$ of the small-scale magnetic field \citep{PFL76},
which describes the dynamical quenching of the $\alpha$ effect.
Note that the dynamical quenching related to evolution of the magnetic
$\alpha$ effect is derived only from the induction equation, and
it is a contribution from small-scale current helicity $\langle {\bm b} {\bf \cdot} (\bec{\nabla} {\bf \times} {\bm b}) \rangle$,
which is related to the small-scale magnetic helicity density.
On the other hand, the algebraic quenching
of the kinetic and magnetic alpha effects and turbulent magnetic diffusion coefficients
of the large-scale magnetic field
are derived from both, the Navier-Stokes equation
for velocity fluctuations and the induction equation for magnetic fluctuations.
In particular, the algebraic quenching is a contribution from the correlation functions for velocity fluctuations
$\langle u_i u_j \rangle$, magnetic fluctuations
$\langle b_i b_j \rangle$ and the cross-helicity correlation function
$\langle u_i b_j \rangle$.
The algebraic quenching is a physical effect related to a feedback of the growing large-scale
magnetic field on plasma motions.
If the algebraic quenching of the $\alpha$ effect is taken into account,
the algebraic quenching of the turbulent magnetic diffusion should be taken into account
as well.
For instance, many studies related to the mean-field numerical simulations
of the evolution of the solar and galactic magnetic fields
take into account algebraic and dynamic quenching
of the $\alpha$ effect, but ignore the algebraic quenching of the turbulent magnetic diffusion
\citep[see, e.g.,][]{CTB97,CTB98,KMR00,KMR02,KKMR03,KSR16,KSR20,KRS23,SKR18,BB02,Brandenburg(2005),SSS06,GCB10,CSS14}.

{\it The approach discussed in this section allows us to derive
the nonlinear turbulent electromotive force
for an intermediate nonlinearity.
This means that the mean magnetic field is not enough strong
to  affect the background turbulence.
The theory for a strong mean magnetic field should  take
into account a modification of the background turbulence by
the mean magnetic field.}

In the next sections we take into account this effect.
In particular, we obtain the dependence of the turbulent kinetic energy
$\meanrho \, \langle {\bm u}^2 \rangle^{(0)}/2$ on the mean magnetic field
using the budget equations for the turbulent kinetic and magnetic energies.
This describes an additional nonlinear effect of
the increase of the turbulent kinetic energy of the background turbulence by
the dissipation of a strong mean magnetic field.
The latter increases turbulent magnetic diffusion
and decreases the nonlinear dynamo number for a strong field,
resulting in a saturation of the dynamo growth of the large-scale magnetic field.

\section{Budget equations}
\label{sect-3}

Using the Navier-Stokes equation for velocity fluctuations, we derive the budget equation
for the density of turbulent kinetic energy (TKE), $E_{_{\rm K}}=\meanrho \, \langle{\bm u}^2\rangle /2$ as
\begin{eqnarray}
{\partial E_{_{\rm K}} \over \partial t} + {\rm div} \, {\bm \Phi}_{_{\rm K}}
= \Pi_{_{\rm K}}  - \varepsilon_{_{\rm K}} ,
\label{T1}
\end{eqnarray}
where ${\bm \Phi}_{_{\rm K}} =\left\langle{\bm u} \left(\rho \, {\bm u}^2 /2  + p\right)\right\rangle - \nu \,\meanrho
\, \bec{\nabla} E_{_{\rm K}}$
is the flux of TKE,
$\varepsilon_{_{\rm K}} = \nu \, \meanrho \, \left\langle \left(\nabla_j u_i\right)^2 \right\rangle$
is the dissipation rate of TKE, and
\begin{eqnarray}
&& \Pi_{_{\rm K}} = - {1 \over \mu_0} \biggl[\left\langle{\bm u} \cdot [{\bm b} \times ({\bm \nabla} \times {\bm b})]\right\rangle
- \left \langle{\bm u} \times (\bec{\nabla} \times {\bm b}) \right\rangle \cdot \meanBB
 \nonumber\\
&& \quad + \left \langle{\bm u} \times {\bm b} \right\rangle \cdot ({\bm \nabla} \times \meanBB) \biggr]
+  \meanrho \,\Big[g \, F_z  - \,\left\langle u_i u_j \right\rangle \, \nabla_j \meanU_i
\nonumber\\
&& \quad + \left\langle {\bm u} \cdot {\bm f} \right\rangle \Big]
\label{T4}
\end{eqnarray}
is the production rate of TKE.
Here $\meanUU$ is the mean velocity,
$\nu$ is the kinematic viscosity and the angular brackets imply ensemble averaging,
${\bm F}=\langle  s \, {\bm u} \rangle$ is the turbulent flux of the entropy,
$s = \theta/\meanT + (\gamma^{-1}-1) p/\meanP $ are entropy fluctuations,
$\theta$ and $\meanT$ are fluctuations and mean fluid temperature,
$\rho$ and $\meanrho$ are fluctuations and mean fluid density,
$p$ and $\meanP$ are fluctuations and mean fluid pressure,
$\gamma=c_{\rm p}/c_{\rm v}$ is the ratio of specific heats,
${\bm g}$  is the acceleration due to the gravity
and $\meanrho \, {\bm f}$ is the external steering force with a zero mean.

We consider three different cases when turbulence is produced
either by convection, or by large-scale shear motions or by an external steering force,
 see the last three terms in the RHS of Eq.~(\ref{T4}).
The first two terms in the RHS of  Eq.~(\ref{T4}) describe an energy exchange between
the turbulent kinetic and magnetic energies (see below), and
the third term in the RHS of Eq.~(\ref{T4}) are due to the work of the Lorentz force
in a nonuniform mean magnetic field.
The estimate for the dissipation rate of the turbulent kinetic energy density
in homogeneous isotropic and incompressible turbulence with a Kolmogorov spectrum is
$\varepsilon_{_{\rm K}} = E_{_{\rm K}} / \tau_0$, where $\tau_0$ is the characteristic
turbulent time at the integral scale.

Using the induction equation for magnetic fluctuations, we derive the budget equation
for the density of turbulent magnetic energy (TME), $E_{_{\rm M}}= \langle{\bm b}^2\rangle /2 \mu_0$ as
\begin{eqnarray}
{\partial E_{_{\rm M}} \over \partial t} + {\rm div} \, {\bm \Phi}_{_{\rm M}}
=\Pi_{_{\rm M}}  - \varepsilon_{_{\rm M}} ,
\label{T5}
\end{eqnarray}
where
\begin{eqnarray}
&& {\bm \Phi}_{_{\rm M}} = {1 \over \mu_0} \biggl[\left\langle{\bm b} \times ({\bm u} \times {\bm b})\right\rangle + \left\langle{\bm u} \, b_j \right\rangle \, \meanB_j -  \left\langle{\bm u} \cdot {\bm b} \right\rangle \, \meanBB
\nonumber\\
&& \quad + \left\langle{\bm b}^2  \right\rangle \, \meanUU - \left\langle{\bm b} \, b_j \right\rangle \, \meanU_j - \eta \, \left \langle{\bm b} \times ({\bm \nabla} \times {\bm b}) \right\rangle \biggr]
\label{T6}
\end{eqnarray}
is the flux of TME,
$\varepsilon_{_{\rm M}} = \eta  \, \left\langle (\bec{\nabla} \times {\bm b})^2 \right\rangle/ \mu_0$ is the dissipation rate of TME, and
\begin{eqnarray}
&& \Pi_{_{\rm M}} = {1 \over \mu_0} \biggl[\left\langle{\bm u} \cdot [{\bm b} \times (\bec{\nabla} \times {\bm b})]\right\rangle
- \left \langle{\bm u} \times ({\bm \nabla} \times {\bm b}) \right\rangle \cdot \meanBB
\nonumber\\
&& \quad + \left\langle b_i \, b_j \right\rangle \, \nabla_j \meanU_i  - \left\langle{\bm b}^2  \right\rangle \,
\left({\bm \nabla} \cdot \meanUU\right) \biggr]
\label{T8}
\end{eqnarray}
is the production rate of TME. Here $\eta$ is the magnetic diffusion due to electrical conductivity of the fluid.
The first two terms in the RHS of  Eq.~(\ref{T8}) describe an energy exchange between
the turbulent magnetic and kinetic energies.
The estimate for the dissipation rate of the turbulent magnetic energy density is
$\varepsilon_{_{\rm M}} = E_{_{\rm M}} / \tau_0$.

The density of total turbulent energy (TTE), $E_{_{\rm T}}=E_{_{\rm K}}+E_{_{\rm M}}$, is determined by the following  budget equation:
\begin{eqnarray}
{\partial E_{_{\rm T}} \over \partial t} + {\rm div} \, {\bm \Phi}_{_{\rm T}}
=\Pi_{_{\rm T}}  - \varepsilon_{_{\rm T}} ,
\label{T5}
\end{eqnarray}
where
\begin{eqnarray}
&& \Pi_{_{\rm T}} = \biggl[\Big(\left\langle b_i \, b_j \right\rangle - \mu_0 \, \meanrho \,\left\langle u_i u_j \right\rangle \Big)\, \nabla_j \meanU_i - \left\langle{\bm b}^2  \right\rangle \, \left({\bm \nabla} \cdot \meanUU\right)
\nonumber\\
&& \; - \left \langle{\bm u} \times {\bm b} \right\rangle \cdot \left({\bm \nabla} \times  \meanBB\right)\biggr] \mu_0^{-1}
+  \meanrho \, \Big(g \, F_z + \left\langle {\bm u} \cdot {\bm f} \right\rangle \Big) .
\label{T9}
\end{eqnarray}
is the production rate of $E_{_{\rm T}}$, $\varepsilon_{_{\rm T}}=\varepsilon_{_{\rm K}}+\varepsilon_{_{\rm M}}$ is the dissipation rate of $E_{_{\rm T}}$ and ${\bm \Phi}_{_{\rm T}}={\bm \Phi}_{_{\rm K}}+{\bm \Phi}_{_{\rm M}}$ is the flux of $E_{_{\rm T}}$.

To determine the production rate of TTE, we use the following second moments for magnetic fluctuations
\citep{RK2007},
\begin{eqnarray}
\left\langle b_i \, b_j \right\rangle = {\meanBB^{\, 2} \over 2} \,   \biggl[2q_{\rm p}\left(\meanB\right) \, \delta_{ij} - q_{\rm s}\left(\meanB\right) \Big(\delta_{ij} + \beta_{ij}\Big) \biggr] ,
\label{C3}
\end{eqnarray}
and velocity fluctuations,
\begin{eqnarray}
 \meanrho \,\left\langle u_i \, u_j \right\rangle &=& - {\meanBB^{\, 2} \over 2\mu_0}   \, \biggl[2q_{\rm p}\left(\meanB\right) \, \delta_{ij} - q_{\rm s}\left(\meanB\right) \Big(\delta_{ij} + \beta_{ij}\Big) \biggr]
\nonumber\\
&& + \meanrho \,\left\langle u_i \, u_j \right\rangle^{(0)}  ,
\label{C5}
\end{eqnarray}
(see Appendix~\ref{sect-A2}),  where  $\beta_{ij}=\meanB_i \meanB_j /\meanBB^{\, 2}$.
The tensor $\left\langle u_i \, u_j \right\rangle^{(0)}$ for a background turbulence (with a zero mean magnetic field)
in Eq.~(\ref{C5}) has two contributions caused by background isotropic velocity fluctuations
 and tangling anisotropic velocity fluctuations due to the mean velocity shear \citep{EKRZ02}:
\begin{eqnarray}
\left\langle u_i \, u_j \right\rangle^{(0)} ={1 \over 3} \left\langle {\bm u}^2\right\rangle^{(0)} \, \delta_{ij}
- 2 \nu_{_{T}}^{(0)} \, \left(\partial \meanU\right)_{ij} ,
\label{C6}
\end{eqnarray}
where $\left(\partial \meanU\right)_{ij} = (\nabla_i \meanU_j + \nabla_j \meanU_i) /2$
and $\nu_{_{T}}^{(0)} = \tau_0 \langle{\bm u}^2  \rangle^{(0)}/3$ is the turbulent viscosity.
For simplicity, in Eq.~(\ref{C3}) we do not take into account a small-scale dynamo with a zero mean magnetic field.

The nonlinear functions $q_{\rm p}(\meanB) $ and $q_{\rm s}(\meanB)$ entering in Eq.~(\ref{C3})--(\ref{C5})
are given by Eqs.~(\ref{X1})--(\ref{X2}) in Appendix~\ref{sect-A2}.
The asymptotic formulae for the nonlinear
functions $q_{\rm p}(\meanB) $ and $q_{\rm s}(\meanB) $ are as follows.
For a very weak mean magnetic field, $ \meanB \ll \meanB_{\rm eq} / 4 {\rm
Rm}^{1/4} $, the nonlinear functions are
given by
\begin{eqnarray}
q_{\rm p}(\meanB)  &=& {2 \over 5} \, \left[\ln {\rm Rm} +
{4 \over 45} \right] ,
\label{T40}\\
q_{\rm s}(\meanB)  &=& {8 \over 15} \,  \left[\ln {\rm Rm} +
{2 \over 15}\right] ,
\label{T41}
\end{eqnarray}
where $\meanB_{\rm eq}^{\, 2} = \mu_0\, \meanrho \, \langle
{\bm u}^2 \rangle$.
For $\meanB_{\rm eq} / 4 {\rm Rm}^{1/4} \ll
\meanB \ll \meanB_{\rm eq} / 4 $, these nonlinear functions are given by
\begin{eqnarray}
q_{\rm p}(\meanB) &=& {16 \over 25} \, \left[5|\ln (\sqrt{2}\beta)| + 1 + 4
\beta^{2}\right] ,
\label{T42}\\
q_{\rm s}(\meanB)  &=& {32 \over 15} \, \left[|\ln (\sqrt{2}\beta)| +
{1 \over 30} + {3 \over 2}  \beta^{2} \right] ,
\label{T43}
\end{eqnarray}
and for $\meanB \gg \meanB_{\rm eq} / 4 $ they are given by
\begin{eqnarray}
q_{\rm p}(\meanB)  &=& {4 \over 3 \beta^2} ,
\quad
q_{\rm s}(\meanB) = {\pi \sqrt{2}\over 3 \beta^3}  .
\label{T45}
\end{eqnarray}
where $\beta = \sqrt{8}  \, \, \meanB/ \meanB_{\rm eq}$.

Substituting Eqs.~(\ref{C3})--(\ref{C6}) into Eq.~(\ref{T9}), we obtain  the production rate of the total turbulent energy as
\begin{eqnarray}
&& \Pi_{_{\rm T}} = \left[ {\meanBB^{\, 2} \over 2 \mu_0} \Big(3q_{\rm p}\left(\meanB\right) - q_{\rm s}\left(\meanB\right) \Big)-
{\meanrho \left\langle {\bm u}^2\right\rangle^{(0)} \over 3} \right] \left({\bm \nabla} \cdot \meanUU\right)
\nonumber\\
&& \;
+ \left[2 \nu_{_{T}}\left(\meanB\right) \, \meanrho \, \left(\partial \meanU\right)_{ij} - {1 \over \mu_0} \, q_{\rm s}\left(\meanB\right)
\, \meanB_i \meanB_j\right] \, \left(\partial \meanU\right)_{ij}
\nonumber\\
&& \;
- {1 \over \mu_0} \,  \bec{\cal E}\left(\meanBB\right) \cdot ({\bm \nabla} \times \meanBB)
+  \meanrho \, \Big(g \, F_z + \left\langle {\bm u} \cdot {\bm f} \right\rangle \Big) ,
\label{T10}
\end{eqnarray}
where $\bec{\cal E}\left(\meanBB\right) = \left \langle{\bm u} \times {\bm b} \right\rangle$ is the turbulent
nonlinear electromotive force.
The turbulent viscosity $\nu_{_{T}}\left(\meanB\right)$ depends on the mean magnetic field.
In particular, for weak field $\meanB \ll \meanB_{\rm eq} / 4$,
the turbulent viscosity $\nu_{_{T}}\left(\meanB\right) \sim \nu_{_{T}}^{(0)} = \tau_0 \langle {\bm u}^2 \rangle^{(0)}/3$,
and for strong field $\meanB \gg \meanB_{\rm eq} / 4$, it is $\nu_{_{T}}\left(\meanB\right) \sim \nu_{_{T}}^{(0)} (1 + \epsilon)/
(4 \meanB / \meanB_{\rm eq})$ \citep{RK2007}.
Using the steady state solution of Eq.~(\ref{T5}), we estimate the total turbulent energy density as
$E_{_{\rm K}}+E_{_{\rm M}} \sim \tau \, \Pi_{_{\rm T}}$, where $\tau$ is of the order of the turbulent time.
Equation~(\ref{C3}) yields the density of turbulent magnetic energy $E_{_{\rm M}}= \langle{\bm b}^2\rangle /2 \mu_0$ as
\begin{eqnarray}
E_{_{\rm M}} = \left[3q_{\rm p}\left(\meanB\right) - 2q_{\rm s} \left(\meanB\right)\right] \, {\meanBB^{\, 2} \over 2\mu_0} .
\label{T11}
\end{eqnarray}
In the next sections, we apply the budget equations for analysis of nonlinear mean-field $\alpha\Omega$, $\alpha^2$
and $\alpha^2\Omega$ dynamos.

\section{Mean-field $\alpha\Omega$ dynamo}
\label{sect-4}

In this section, we consider the axisymmetric mean-field $\alpha\Omega$ dynamo, so that the mean
magnetic field can be decomposed as
\begin{eqnarray}
\meanBB=\meanB_{y}(t,x,z) {\bm e}_{y} + {\rm rot} [\meanA(t,x,z) {\bm e}_{y}],
\label{QS1}
\end{eqnarray}
and nonlinear mean-field induction equation reads
\begin{eqnarray}
{\partial  \over \partial t}
\begin{pmatrix}
\meanA \\ \meanB_{y}
\end{pmatrix}
= {\hat N}
\begin{pmatrix}
\meanA \\ \meanB_{y}
\end{pmatrix} ,
\label{QS2}
\end{eqnarray}
where the operator ${\hat N}$ is given by
\begin{eqnarray}
{\hat N} &=&
\begin{pmatrix}
\eta_{_{T}}^{(A)}\left(\meanB\right) \Delta & \quad \alpha\left(\meanBB\right) \\
\\
R_{\alpha} R_\omega \, \hat \Omega  & \quad \nabla_j \, \eta_{_{T}}^{(B)}\left(\meanB\right) \nabla_j
\end{pmatrix} ,
\label{QS3}
\end{eqnarray}
and the operator
\begin{eqnarray}
\hat \Omega \, \meanA= {\partial(\delta \Omega \, \sin \vartheta, \, \meanA) \over \partial(z, \, x)}
\label{QS4}
\end{eqnarray}
describes differential rotation.
Here $\vartheta$ is the angle between $\delta {\bm \Omega}$ and the vertical coordinate $z$ and $L$
is the characteristic scale (e.g., the radius of a star or the thickness of a galactic disk).
The total $\alpha$ effect is the sum of the kinetic $\alpha$ effect, $\alpha_{_{\rm K}}(\meanB)$, and
the magnetic $\alpha$ effect, $\alpha_{_{\rm M}}(\meanBB)$,
$\alpha(\meanBB)=\alpha_{_{\rm K}}(\meanB) + \alpha_{_{\rm M}}(\meanBB)$,
where the kinetic $\alpha$ effect is proportional to
the kinetic helicity $H_{\rm u} = \langle {\bm u} {\bf \cdot} (\bec{\nabla} {\bf \times} {\bm u}) \rangle$,
and the magnetic $\alpha$ effect is proportional to the current helicity
$H_{\rm c}\left(\meanBB\right)=\langle {\bm b} {\bf \cdot} (\bec{\nabla} {\bf \times} {\bm b}) \rangle$
of the small-scale magnetic field ${\bm b}$.

Equations~(\ref{QS2})--(\ref{QS4}) are written in dimensionless variables:
the coordinate is measured in the units of $L$, the time $t$ is
measured in the units of turbulent magnetic diffusion time $L^2 / \eta_{_{T}}^{(0)}$;
the mean magnetic field is measured in the units of $\meanB_\ast$, where
$\meanB_\ast \equiv  \sigma \,\, \meanB_\ast^{\,\rm eq}$,
$\sigma= \ell_0/\sqrt{2}L$, $\meanB_\ast^{\,\rm eq} = u_0 \, \sqrt{\mu_0 \meanrho_\ast}$,
and the  magnetic potential, $\meanA$ is measured in the units of
$R_{\alpha} L \meanB_\ast$. Here $R_{\alpha} = \alpha_\ast L / \eta_{_{T}}^{(0)}$,
the fluid density $\meanrho$ is measured in the units $\meanrho_\ast$,
the differential rotation $\delta\Omega$ is measured in units of the maximal value
of the angular velocity $\Omega$,
the $\alpha$ effect is measured in units of the maximum value of the
kinetic $ \alpha $ effect, $\alpha_\ast$;
the integral scale of the turbulent motions
$\ell_0=\tau_0 \, u_0$ and the characteristic turbulent velocity
$u_0=\sqrt{\langle{\bm u}^2  \rangle^{(0)}}$
at the scale $\ell_0$ are measured in units of their
maximum values in the turbulent region, and the turbulent magnetic diffusion coefficients
are measured in units of their maximum values.
The magnetic Reynolds number ${\rm Rm}=\ell_0 \, u_0/\eta$
is defined using the maximal values of the integral scale $\ell_0$ and
the characteristic turbulent velocity $u_0$.
The dynamo number for the linear $\alpha\Omega$ dynamo is defined as $D_{\rm L} = R_\alpha
R_\omega$, where $R_\omega = (\delta \Omega) \, L^2 / \eta_{_{T}}^{(0)}$.

Now we define the nonlinear dynamo number $D_{\rm N}\left(\meanB\right)$
for the $\alpha\Omega$ dynamo as
\begin{eqnarray}
D_{\rm N}\left(\meanB\right) = {\alpha\left(\meanB\right) \, \delta \Omega \, L^3 \over \eta_{_{T}}^{(B)}\left(\meanB\right) \, \eta_{_{T}}^{(A)}\left(\meanB\right)},
\label{RS1}
\end{eqnarray}
where we take into account that
the nonlinear turbulent magnetic diffusion coefficients of the poloidal and toroidal components of the mean magnetic field
are different \citep{RK2004}.

Next, we take into account the feedback of the mean magnetic field on the background turbulence
using the budget equation for the total turbulent energy.
In a shear-produced non-convective turbulence, the leading-order contributions to the production rate of
the turbulent kinetic energy for a strong large-scale magnetic field ($\meanB \gg \meanB_{\rm eq} / 4$)
is due to the term $- \bec{\cal E}\left(\meanBB\right) \cdot ({\bm \nabla} \times \meanBB) /\mu_0$, so that
the leading-order contribution to the turbulent kinetic energy density for a strong large-scale magnetic field
is estimated as
\begin{eqnarray}
E_{_{\rm K}} = -{\tau  \over \mu_0} \, \bec{\cal E}\left(\meanBB\right) \cdot ({\bm \nabla} \times \meanBB) .
\label{TT18}
\end{eqnarray}
Indeed, let us estimate the leading-order contributions to the production rate of the total turbulent energy given by ~(\ref{T10}).
Using Eqs.~(\ref{WEF11})--(\ref{WEF12}), we can rewrite
the turbulent electromotive force as ${\cal E}_i = \alpha \meanB_i - \eta_{ij}^{\rm (T)} ({\bm \nabla} \times \meanBB)_j$,
where $\eta_{ij}^{\rm (T)}$ is the diagonal tensor
with components $\eta_{11}^{\rm (T)}=\eta_{_{T}}^{(A)}$ and
$\eta_{22}^{\rm (T)}=\eta_{_{T}}^{(B)}$.
Now we estimate:
\begin{eqnarray*}
\eta_{ij}^{\rm (T)} ({\bm \nabla} \times \meanBB)_j ({\bm \nabla} \times \meanBB)_i =
\eta_{_{T}}^{(A)} \, ({\bm \nabla} \times \meanBB)^2_\varphi +
\eta_{_{T}}^{(B)} \, ({\bm \nabla} \times \meanBB)^2_{\rm p} ,
\end{eqnarray*}
where $({\bm \nabla} \times \meanBB)_\varphi$ and
$({\bm \nabla} \times \meanBB)_{\rm p}$ are
the toroidal and poloidal components of the electric current, which can be estimated as:
$|({\bm \nabla} \times \meanBB)_\varphi| \sim |\meanB_{\rm p}|/L_B$
and $|({\bm \nabla} \times \meanBB)_{\rm p}| \sim |\meanB_\varphi|/L_B$.
Here the characteristic scale of the mean magnetic field variations $L_B$
is defined as $L_B = \meanB /|{\bm \nabla} \times \meanBB|$.
We also take into account that for a strong field ($\meanB \gg \meanB_{\rm eq} / 4$),
$\eta_{_{T}}^{(A)}(\beta) \sim  \eta_{_{T}}^{(0)}  /\beta^2$, while $\eta_{_{T}}^{(B)}(\beta) \sim  \eta_{_{T}}^{(0)} / \beta$,
where $\meanB_\varphi$ and $\meanB_{\rm p}$ are the toroidal and poloidal components of the mean magnetic field.
For the $\alpha\Omega$ dynamo,  the toroidal component of the mean magnetic field is much larger than
the  poloidal component, i.e., $|\meanB_{\rm p}| \ll |\meanB_\varphi|$.
This yields
\begin{eqnarray}
- \bec{\cal E}\left(\meanBB\right) \cdot ({\bm \nabla} \times \meanBB)
\sim {\eta_{_{T}}^{(B)} \over L_B^2}\, \meanB_\varphi^{\, 2} \sim {\eta_{_{T}}^{(0)} \over 4L_B^2}
\,\meanB_\varphi \, \meanB_{\rm eq} ,
\label{QT18}
\end{eqnarray}
where the magnetic energy of the equipartition field $\meanB_{\rm eq}$ is defined as $\meanB_{\rm eq}^2/2\mu_0
= E_{_{\rm K}}^{(0)}$. For a shear-produced turbulence $E_{_{\rm K}}^{(0)} \approx \, \meanrho \, \ell_0^2 \, S^2$
with the squared shear $S^2=\left(\partial \meanU\right)_{ij}^2$
and $\ell_0 = \tau [\langle {\bm u}^2 \rangle^{(0)}]^{1/2}$ being the integral scale of turbulence at vanishing
mean magnetic field. We assume also that the correlation time is independent of the mean magnetic field.

Contributions of other terms to the production rate of TTE and TKE
for a strong large-scale magnetic field are much smaller than that described by Eq.~(\ref{QT18}).
For instance,  the contribution $\alpha \meanBB \cdot ({\bm \nabla} \times \meanBB)$
to $- \bec{\cal E}\left(\meanBB\right) \cdot ({\bm \nabla} \times \meanBB)$ is much smaller for a strong field, because
\begin{eqnarray*}
\meanBB \cdot ({\bm \nabla} \times \meanBB) = \meanB_{\rm p} \,  ({\bm \nabla} \times \meanBB)_{\rm p} +
\meanB_\varphi \,  ({\bm \nabla} \times \meanBB)_\varphi \sim {\meanB_{\rm p} \meanB_\varphi \over L_B} ,
\end{eqnarray*}
and for a strong field $\alpha(\beta) \sim \alpha^{(0)} / \beta^{2}$.
Similarly, the checking of the contributions of the remaining terms to the production rate of TTE and TKE
for a strong large-scale magnetic field shows that they are much smaller than that described by Eq.~(\ref{QT18}).
Therefore, the leading-order contribution to the turbulent kinetic energy density
$E_{_{\rm K}}\left(\meanB\right)$ for strong mean magnetic fields is
\begin{eqnarray}
E_{_{\rm K}}\left(\meanB\right) \sim {E_{_{\rm K}}^{(0)} \over 6} \,  \left({\ell_0 \over L_B} \right)^2 \left({\meanB \over \meanB_{\rm eq}} \right) .
\label{T19}
\end{eqnarray}
Equation~(\ref{T19}) implies that the turbulent kinetic energy increases due to the dissipation of the strong large-scale magnetic field.

This yields the estimate for the turbulent magnetic diffusion coefficient of toroidal magnetic field $\eta_{_{T}}^{(B)}\left(\meanB\right) = \eta_{_{T}}^{(0)} \, \phi_\eta^{(B)} \, E_{_{\rm K}}\left(\meanB\right) / E_{_{\rm K}}^{(0)}$ in the limit of a strong field  as
\begin{eqnarray}
{\eta_{_{T}}^{(B)}\left(\meanB\right) \over \eta_{_{T}}^{(0)}} &\approx&{1 \over 24}  \left({\ell_0 \over L_B} \right)^2 = const ,
\label{T16}
\end{eqnarray}
where $\eta_{_{T}}^{(0)}=2  \tau \, E_{_{\rm K}}^{(0)}/ 3 \meanrho$ and we take into account the increase of
the turbulent kinetic energy caused by the dissipation of the strong large-scale magnetic field [see Eq.~(\ref{T19})].
As follows from Eq.~(\ref{PB7}),
the ratio of turbulent magnetic diffusion coefficients of poloidal and toroidal fields $\eta_{_{T}}^{(A)}\left(\meanB\right)/\eta_{_{T}}^{(B)}\left(\meanB\right)$ is given by
\begin{eqnarray}
{\eta_{_{T}}^{(A)}\left(\meanB\right) \over \eta_{_{T}}^{(B)}\left(\meanB\right)} \approx {1 \over 2}\left({\meanB \over \meanB_{\rm eq}} \right)^{-1} .
\label{AT16}
\end{eqnarray}

The dependence of the total $\alpha$ effect on the mean magnetic field, $\alpha\left(\meanB\right)$, is caused by
the  algebraic and dynamic quenching.
The algebraic quenching describes the feedback
of the mean magnetic field on the plasma motions, while
the dynamic quenching of the total $\alpha$ effect is
caused by the evolution of the magnetic $\alpha$ effect
related to the small-scale current and magnetic helicities.
In particular, the dynamic equation
for the small-scale current helicity (which determines the
evolution of the magnetic $\alpha$ effect) in a steady state yields the estimate for the total $\alpha$ effect
in the limit of a strong mean field as
$\alpha\left(\meanB\right) \propto - {\rm div} {\bm F}_{_{\rm M}} / \meanB^2$, where
${\bm F}_{_{\rm M}}$ is the magnetic helicity flux of the small-scale magnetic field.
This implies that the total $\alpha$ effect for strong magnetic fields behaves as
\begin{eqnarray}
{\alpha\left(\meanB\right)  \over \alpha_{_{\rm K}}^{(0)}} \propto \left({\meanB  \over \meanB_{\rm eq}} \right)^{-2} .
\label{UT16}
\end{eqnarray}
Note that the algebraic and dynamic quenching of the alpha effect
yield similar behavior for a strong large-scale magnetic field
[see Eqs.~(\ref{PB5})--(\ref{PB6}) and~(\ref{UT16}), and paper by~\cite{CSS14}].

Therefore, the ratio $D_{\rm N}\left(\meanB\right)/D_{\rm L}$ of the nonlinear and linear dynamo numbers
in a shear-produced turbulence for strong mean magnetic fields
is estimated as [see Eqs.~(\ref{RS1}) and (\ref{T16})--(\ref{UT16})]:
\begin{eqnarray}
{D_{\rm N}\left(\meanB\right)\over D_{\rm L}}  &\approx& 2 \left({\meanB \over \meanB_{\rm eq}} \right)^{-1}  \, \left({\eta_{_{T}}^{(B)}\over \eta_{_{T}}^{(0)}}\right)^{-2} \propto  \left({\meanB \over \meanB_{\rm eq}} \right)^{-1} .
\label{T17}
\end{eqnarray}
Equation~(\ref{T17}) implies that the
nonlinear dynamo number decreases with the increase of the mean magnetic field
for any strong values of the field for a shear-produced turbulence.
This results in saturation of the mean-field dynamo instability.

In a convective turbulence,  the largest contributions to the production rate of TTE
for a strong mean magnetic fields
is due to the buoyancy term $\meanrho \, g \, F_z$ and the term
$\eta_{_{T}}^{(B)}\left(\meanB\right) \, ({\bm \nabla} \times \meanBB)^2 /\mu_0$ [see Eq.~(\ref{T10})].
This implies that the leading-order contribution to the turbulent kinetic energy density
$E_{_{\rm K}}\left(\meanB\right)$ in a convective turbulence for strong mean magnetic fields is given by
Eq.~(\ref{T19}), where $E_{_{\rm K}}^{(0)} = (\meanrho / 2) \, (2 g \, F_z \, \ell_0)^{2/3}$.
Therefore, equations  for the ratios $\eta_{_{T}}^{(B)}\left(\meanB\right)/\eta_{_{T}}^{(0)}$,
$\eta_{_{T}}^{(A)}\left(\meanB\right)/\eta_{_{T}}^{(B)}\left(\meanB\right)$
and $D_{\rm N}\left(\meanB\right)/D_{\rm L}$  in a convective turbulence for strong mean magnetic fields
are similar to Eqs.~(\ref{T16})--(\ref{T17}), respectively.
The difference is only in equation for $E_{_{\rm K}}^{(0)}$  that  for a convective turbulence is given by
$E_{_{\rm K}}^{(0)} = (\meanrho / 2) \, (2 g \, F_z \, \ell_0)^{2/3}$
and for a shear-produced turbulence
is $E_{_{\rm K}}^{(0)} = (2/3) \, \meanrho \, \ell_0^2 \, S^2$.
The similar situation is also for a forced turbulence except for
the expression for $E_{_{\rm K}}^{(0)}$ for a forced turbulence
reads $E_{_{\rm K}}^{(0)} = \meanrho \, \tau_0 \, \left\langle {\bm u} \cdot {\bm f} \right\rangle$.

This implies that for the $\alpha\Omega$ dynamo, the nonlinear
dynamo number decreases with increase of
the mean magnetic field for a forced turbulence, and a shear-produced turbulence
and a convective turbulence. This causes saturation of the mean-field $\alpha\Omega$ dynamo instability
for a strong mean magnetic field.

\section{Mean-field $\alpha^2$ dynamo}
\label{sect-5}

In this section, we consider mean-field $\alpha^2$ dynamo.
First, we discuss a long-standing question: ``When can a one-dimensional kinematic
$\alpha^2$ dynamo be oscillatory?"
The mean magnetic field $\meanBB(t,z)={\bm \nabla} \times \meanAA =
(-\nabla_z \meanA_y, \nabla_z \meanA_x, 0)$
is determined by the following equation
\begin{eqnarray}
{\partial \Psi \over \partial t} = {\hat L} \Psi ,
\label{A1}
\end{eqnarray}
where $\meanAA$ is the mean magnetic vector potential in the Weyl
gauge. The linear operator ${\hat L}$ and the function $\Psi(t,z)$ are given by
\begin{eqnarray}
{\hat L} &=&
\begin{pmatrix}
\eta_{_{T}}^{(0)} \nabla_z^2 & \quad - \alpha_{_{\rm K}}^{(0)} \nabla_z \\
\alpha_{_{\rm K}}^{(0)} \nabla_z  & \quad \eta_{_{T}}^{(0)} \nabla_z^2
\end{pmatrix} , \quad
\Psi =
\begin{pmatrix}
A_x \\ A_y
\end{pmatrix} ,
\label{A3}
\end{eqnarray}
where $\eta_{_{T}}^{(0)}$ is the turbulent magnetic diffusion coefficient, and $\alpha_{_{\rm K}}^{(0)}$ is the kinetic
$\alpha$ effect caused by the helical turbulent motions in plasma.
If the linear operator ${\hat L}$ is not self-adjoint, it has
complex eigenvalues. This case corresponds to the oscillatory growing solution, i.e.,
the dynamo is oscillatory.
On the other hand, any self-adjoint operator, ${\hat M}$, defining by the
following condition,
\begin{eqnarray}
\int \Psi^\ast {\hat M} \tilde \Psi \,dz = \int \tilde \Psi {\hat M}^\ast \Psi^\ast \,dz ,
\label{A4}
\end{eqnarray}
has real eigenvalues, where the asterisk denotes complex conjugation.
Now we determine conditions when
the linear operator ${\hat L}$ is not self-adjoint, i.e., it has
complex eigenvalues.
To this end, we determine the integrals $\int \Psi^\ast {\hat L} \tilde \Psi \,dz$
and $\int \tilde \Psi {\hat L}^\ast \Psi^\ast \,dz$ as:
\begin{eqnarray}
&& \int \Psi^\ast {\hat L} \tilde \Psi \,dz = \int \alpha_{_{\rm K}}^{(0)} \left(A_y^\ast \nabla_z \tilde A_x
- A_x^\ast \nabla_z \tilde A_y\right) \,d z
\nonumber\\
&& \quad \quad- \int \eta_{_{T}}^{(0)} \left[\left(\nabla_z A_x^\ast\right) \, \nabla_z \tilde A_x
+ \left(\nabla_z A_y^\ast\right) \, \nabla_z \tilde A_y \right] \,d z
\nonumber\\
&& \quad \quad+ \left[\eta_{_{T}}^{(0)} \left(A_x^\ast \, \nabla_z \tilde A_x
+ A_y^\ast \, \nabla_z \tilde A_y\right) \right]_{z=L_{\rm bott}}^{z=L_{\rm top}},
\label{A5}
\end{eqnarray}

\begin{eqnarray}
&& \int \tilde \Psi {\hat L}^\ast \Psi^\ast \,dz = \int \alpha_{_{\rm K}}^{(0)} \left(A_y^\ast \nabla_z \tilde A_x
- A_x^\ast \nabla_z \tilde A_y\right) \,d z
\nonumber\\
&& \quad \quad- \int \eta_{_{T}}^{(0)} \left[\left(\nabla_z A_x^\ast\right) \, \nabla_z \tilde A_x
+ \left(\nabla_z A_y^\ast\right) \, \nabla_z \tilde A_y \right] \,d z
\nonumber\\
&& \quad \quad+ \biggl[\eta_{_{T}}^{(0)} \biggl(\tilde A_x \, \nabla_z A_x^\ast
+ \tilde A_y \, \nabla_z A_y^\ast \biggr) + \alpha_{\rm k} \biggl(A_x^\ast \tilde A_y
\nonumber\\
&& \quad \quad - A_y^\ast \, \tilde A_x\biggr)\biggr]_{z=L_{\rm bott}}^{z=L_{\rm top}} \; ,
\label{A6}
\end{eqnarray}
where $z=L_{\rm bott}$ and $z=L_{\rm top}$ are the bottom and upper boundaries, respectively.
When $\eta_{_{T}}^{(0)}$ and $\alpha_{_{\rm K}}^{(0)}$ vanish at the boundaries
where the turbulence is very weak, the operator ${\hat L}$ satisfies condition~(\ref{A4})
and the $\alpha^2$ dynamo is not oscillatory.
On the other hand, when $\alpha_{_{\rm K}}^{(0)}$ vanishes only at one boundary, while
it is non-zero at the other boundary, the operator ${\hat L}$ does not satisfy
condition~(\ref{A4}), and the $\alpha^2$ dynamo is oscillatory.
The latter case has been considered in analytical study by \cite{SSR85,RB87} and
in numerical study by \cite{BS87}.
\cite{B17} has recently considered the one-dimensional kinematic $\alpha^2$ dynamo
with different conditions at two boundaries: ${\bm A}=0$ at $z=L_{\rm bott}$ and $\nabla_z{\bm A}=0$
at $z=L_{\rm top}$, so that the operator ${\hat L}$ may not satisfy
condition~(\ref{A4}), and the $\alpha^2$ dynamo may be oscillatory.

Now we consider the nonlinear axisymmetric mean-field $\alpha^2$ dynamo, so that
nonlinear mean-field induction equation reads
\begin{eqnarray}
{\partial  \over \partial t}
\begin{pmatrix}
\meanA \\ \meanB_y
\end{pmatrix}
= {\hat N}
\begin{pmatrix}
\meanA \\ \meanB_y
\end{pmatrix} ,
\label{L2}
\end{eqnarray}
where the mean magnetic field is $\meanBB=\meanB_{y}(t,x,z) {\bm e}_{y} + {\rm rot} [\meanA(t,x,z) {\bm e}_{y}]$,
the operator ${\hat N}$ is given by
\begin{eqnarray}
{\hat N} &=&
\begin{pmatrix}
\eta_{_{T}}^{(A)}\left(\meanB\right) \Delta & \quad \alpha\left(\meanB\right)\\
\\
- R_{\alpha}^2 \nabla_j \alpha\left(\meanB\right) \nabla_j  & \quad \nabla_j \eta_{_{T}}^{(B)}\left(\meanB\right) \nabla_j
\end{pmatrix} ,
\label{L3}
\end{eqnarray}
and the total $\alpha$ effect is given by $\alpha\left(\meanB\right) =\alpha_{_{\rm K}}\left(\meanB\right) + \alpha_{_{\rm M}}\left(\meanBB\right)$.
Now we introduce the effective dynamo number $D_{\rm N}^{(\alpha)}\left(\meanB\right)$
in the nonlinear $\alpha^2$ dynamo defined as
$D_{\rm N}^{(\alpha)}\left(\meanB\right) = \alpha^2\left(\meanB\right) \, L^2/ [\eta_{_{T}}^{(B)}\left(\meanB\right) \eta_{_{T}}^{(A)}\left(\meanB\right)]$.
Similarly,  the effective dynamo number for a linear $\alpha^2$ dynamo is defined as
$D_{\rm L}^{(\alpha)}=R_{\alpha}^2$,
where $R_{\alpha} = \alpha_\ast L / \eta_{_{T}}^{(0)}$, $\alpha_\ast$ is the maximum value of the kinetic $\alpha$ effect and $L$ is the stellar radius or the thickness of the galactic disk.

The poloidal and toroidal components of the mean magnetic field in the nonlinear $\alpha^2$ mean-field dynamo
are of the same order of magnitude.
Equations~ (\ref{T19})--(\ref{UT16}) obtained in Section~\ref{sect-4}
can be used for the nonlinear $\alpha^2$ mean-field dynamo as well.
Therefore, the ratio $D_{\rm N}^{(\alpha)}\left(\meanB\right)/D_{\rm L}^{(\alpha)}$ for strong mean magnetic fields is given by
\begin{eqnarray}
{D_{\rm N}^{(\alpha)}\over D_{\rm L}^{(\alpha)}} &\approx& \left({\meanB \over \meanB_{\rm eq}} \right)^{-3} .
\label{L4}
\end{eqnarray}
These equations take into account the feedback
of the mean magnetic field on the background turbulence
by means of the budget equation for the total turbulent energy.
Thus, Eq.~(\ref{L4}) implies that for the $\alpha^2$ dynamo, the nonlinear
dynamo number decreases with increase of
the mean magnetic field.
This causes a saturation of the mean-field $\alpha^2$ dynamo instability
for a strong mean magnetic field.

\section{Mean-field $\alpha^2\Omega$ dynamo}
\label{sect-6}

In this section, we consider the axisymmetric mean-field $\alpha^2\Omega$ dynamo, so that
and nonlinear mean-field induction equation reads
\begin{eqnarray}
{\partial  \over \partial t}
\begin{pmatrix}
\meanA \\ \meanB_{y}
\end{pmatrix}
= {\hat N}
\begin{pmatrix}
\meanA \\ \meanB_{y}
\end{pmatrix} ,
\label{S2}
\end{eqnarray}
where the mean magnetic field is $\meanBB=\meanB_{y}(t,x,z) {\bm e}_{y} + {\rm rot} [\meanA(t,x,z) {\bm e}_{y}]$,
the operator ${\hat N}$ is
\begin{eqnarray}
{\hat N} &=&
\begin{pmatrix}
\eta_{_{T}}^{(A)}\left(\meanB\right) \Delta & \alpha\left(\meanBB\right) \\
\\
R_{\alpha} \left[R_\omega \hat \Omega - R_{\alpha} \nabla_j \alpha\left(\meanB\right) \nabla_j\right]   &  \nabla_j \eta_{_{T}}^{(B)}\left(\meanB\right) \nabla_j
\end{pmatrix} ,
\nonumber\\
\label{S3}
\end{eqnarray}
and $R_{\alpha} = \alpha_\ast L / \eta_{_{T}}^{(0)}$ and $R_\omega = (\delta \Omega) \, L^2 / \eta_{_{T}}^{(0)}$.
The kinematic and weakly nonlinear $\alpha^2 \, \Omega$ dynamos have been studied using asymptotic analysis
\citep{MNS96,GSK01,BSK05}.

We consider a kinematic dynamo problem,
assuming for simplicity that the kinetic $\alpha$ effect
is a constant, and the mean velocity $\overline{\bm U} = (0, Sz, 0)$, where $S \equiv \delta \Omega$.
We seek a solution for the linearised equation~(\ref{S2}) as a real part of the following functions:
\begin{eqnarray}
\meanA=A_0 \exp[\tilde \gamma t - {\rm i} \, (k_x x+ k_z z)],
\label{TM3}
\end{eqnarray}
\begin{eqnarray}
\meanB_\varphi = B_0 \exp[\tilde \gamma t - {\rm i} \, (k_x x+ k_z z)],
\label{TM4}
\end{eqnarray}
where $\tilde \gamma=\gamma + {\rm i} \, \omega$.
Equations~(\ref{S2})--(\ref{TM4}) yield the growth rate of the dynamo instability and the frequency of the dynamo waves as
\begin{eqnarray}
\gamma &=& {R_{\alpha} R_{\alpha}^{\rm cr} \over \sqrt{2}} \left[\left[1 + \left({\zeta R_\omega \over R_{\alpha} R_{\alpha}^{\rm cr}}\right)^2\right]^{1/2} + 1 \right]^{1/2}
- \left(R_{\alpha}^{\rm cr}\right)^2 ,
\nonumber\\
\label{TM5}
\end{eqnarray}
\begin{eqnarray}
\omega = - {\rm sgn}(R_\omega) \, {R_{\alpha} R_{\alpha}^{\rm cr} \over \sqrt{2}} \left[\left[1 + \left({\zeta R_\omega \over R_{\alpha} R_{\alpha}^{\rm cr}}\right)^2\right]^{1/2} - 1 \right]^{1/2}  ,
\label{TM6}
\end{eqnarray}
where $\zeta^2=1 - \left(k_x/R_{\alpha}^{\rm cr}\right)^2$.
Here we took into account that
$(x + {\rm i}y)^{1/2}= \pm (X + {\rm i}Y)$, where
$X = 2^{-1/2} \, [(x^2+y^2)^{1/2} + x]^{1/2}$ and $Y = {\rm sgn}(y) \, 2^{-1/2} \, [(x^2+y^2)^{1/2} - x]^{1/2}$.
Here the threshold $R_{\alpha}^{\rm cr}$ for the mean-field dynamo instability, defined by the conditions
$\gamma=0$ and $R_\omega=0$, is given by $R_{\alpha}^{\rm cr}=(k_x^2 + k_z^2)^{1/2}$.

Equations~(\ref{S2})--(\ref{TM4}) also yield the squared ratio of amplitudes $|A_0 / B_0|^2$,
\begin{eqnarray}
\left|{A_0 \over B_0}\right|^2 =  \left(R_{\alpha} R_{\alpha}^{\rm cr} \right)^{-2} \,
\left[1 + \left({\zeta R_\omega \over R_{\alpha} R_{\alpha}^{\rm cr}}\right)^2\right]^{-1/2} ,
\label{TM11}
\end{eqnarray}
and the phase shift $\delta$ between the toroidal field $\overline{B}_\varphi$ and the magnetic vector potential $\overline{A}$ is given by
\begin{eqnarray}
\sin(2\delta) =  - \zeta R_\omega \,  \left[\left(R_{\alpha} R_{\alpha}^{\rm cr} \right)^{2} + \zeta^2 R_\omega^2\right]^{-1/2} .
\label{TM14}
\end{eqnarray}
Equation~(\ref{TM11}) yields the energy ratio of poloidal $\meanB_{\rm pol}$ and toroidal $\meanB_\varphi$ mean magnetic field components as
\begin{eqnarray}
{\meanB_{\rm pol}^2 \over \meanB_\varphi^2} =  \left[1 + \left({\zeta R_\omega \over R_{\alpha} R_{\alpha}^{\rm cr}}\right)^2\right]^{-1/2} ,
\label{TM12}
\end{eqnarray}
where $\meanB_{\rm pol}^2 = \meanB_x^2 + \meanB_z^2= (R_{\alpha} R_{\alpha}^{\rm cr} \, \meanA)^2$.

Asymptotic formulas for the growth rate of the dynamo instability and the frequency of the dynamo waves for a weak differential rotation, $\zeta R_\omega \ll R_{\alpha} R_{\alpha}^{\rm cr}$,  are given by
\begin{eqnarray}
\gamma = R_{\alpha} R_{\alpha}^{\rm cr} \left[1 + {1 \over 8} \left({\zeta R_\omega \over R_{\alpha} R_{\alpha}^{\rm cr}}\right)^2\right] - \left(R_{\alpha}^{\rm cr}\right)^2 ,
\label{TM7}
\end{eqnarray}
\begin{eqnarray}
\omega =- {\zeta R_\omega \over \sqrt{2}} .
\label{TM8}
\end{eqnarray}
In this case, the mean-field $\alpha^2$ dynamo is slightly modified by a weak differential rotation, and
the phase shift between the fields $\overline{B}_\varphi$ and $\overline{B}_{\rm pol}$ vanishes,
while $\overline{B}_{\rm pol}/  \overline{B}_\varphi \sim 1$ [see Eqs.~(\ref{TM14})--(\ref{TM12})].
In the opposite case, for a strong differential rotation, $\zeta R_\omega \gg R_{\alpha} R_{\alpha}^{\rm cr}$, the growth rate of the dynamo instability and the frequency of the dynamo waves are given by
\begin{eqnarray}
\gamma = \left[ {1 \over 2} \,  \zeta \, R_{\alpha}^{\rm cr} \, R_{\alpha} |R_\omega| \right]^{1/2} - \left(R_{\alpha}^{\rm cr}\right)^2 ,
\label{TM9}
\end{eqnarray}
\begin{eqnarray}
\omega = - {\rm sgn}(R_\omega) \left[ {1 \over 2} \,  \zeta \, R_{\alpha}^{\rm cr} \, R_{\alpha} |R_\omega| \right]^{1/2} .
\label{TM10}
\end{eqnarray}
In this case, the mean-field $\alpha\Omega$ dynamo is slightly modified by a weak $\alpha^2$ effect,
and the phase shift between the fields $\overline{B}_\varphi$ and $\overline{B}_{\rm pol}$ tends to $- \pi/4$,
while $\overline{B}_{\rm pol}/  \overline{B}_\varphi \ll 1$ [see Eqs.~(\ref{TM14})--(\ref{TM12})].
The necessary condition for the dynamo ($\gamma>0$) in this case reads:
\begin{itemize}
\item{
when $R_{\alpha}/R_{\alpha}^{\rm cr} < \sqrt{2}$, the mean-field $\alpha^2 \, \Omega$ dynamo
is excited when
\begin{eqnarray}
D_{\rm L} > {2 \over \zeta} \, \left(R_{\alpha}^{\rm cr}\right)^3 ;
\label{TL1}
\end{eqnarray}
}
\item{
when $R_{\alpha}/R_{\alpha}^{\rm cr} > \sqrt{2}$, the mean-field $\alpha^2 \, \Omega$ dynamo
is excited for any differential rotation, $R_\omega$. Here $D_{\rm L}=R_{\alpha} \, R_\omega$.}
\end{itemize}

Analysis which is similar to that performed in Section~\ref{sect-4}
[see Eqs.~ (\ref{T19})--(\ref{UT16})]
yields the ratio of the nonlinear and linear dynamo numbers $D_{\rm N}\left(\meanB\right)/D_{\rm L}$
in the nonlinear $\alpha^2\Omega$ dynamo for strong mean magnetic fields
that is coincided with Eq.~ (\ref{L4}).
The latter implies that for the $\alpha^2\Omega$ dynamo, the nonlinear
dynamo number decreases with increase of
the mean magnetic field, so that the nonlinear mean-field  dynamo  instability
is always saturated for strong mean magnetic fields.

\section{Conclusions}
\label{sect-7}

In the sun, stars and galaxies, the large-scale magnetic fields are originated due to
the mean-field dynamo instabilities.
The saturation of the dynamo generated large-scale magnetic fields
is caused by algebraic and dynamic nonlinearities.
A key parameter which controls the  saturation of the $\alpha\Omega$ dynamo
instability is the nonlinear dynamo number
$D_{\rm N}\left(\meanB\right) = \alpha\left(\meanB\right) \,
\delta \Omega \, L^3/ \eta_{_{T}}^2\left(\meanB\right)$.
When the total $\alpha$ effect and the turbulent magnetic diffusion are quenched as
$(\meanB/ \meanB_{\rm eq})^{-2}$ for strong mean magnetic fields,
the nonlinear dynamo number $D_{\rm N}\left(\meanB\right)$
increases with the increase of the large-scale magnetic field.
The latter implies that the mean-field dynamo instability
cannot be saturated for a strong field.

In the present study we have shown that the dissipation of the generated strong
large-scale magnetic field increases both, the turbulent kinetic energy
of the background turbulence and the turbulent magnetic diffusion
coefficient.
This additional nonlinear effect decreases the nonlinear dynamo number for a strong field
and causes a saturation of the dynamo growth of
large-scale magnetic field.
This nonlinear effect is taken into account by means
of the budget equation for the total turbulent energy.
Using this approach and considering various origins of turbulence
(e.g., a forced turbulence, a shear-produced turbulence
and a convective turbulence), we  have demonstrated that
the mean-field $\alpha\Omega$, $\alpha^2$ and $\alpha^2\Omega$ dynamo instabilities
can be always saturated for any strong mean magnetic field.
These results have very important applications for astrophysical magnetic fields.

\section*{Acknowledgments}

The detailed comments on our manuscript by the
anonymous referee which essentially improved the presentation
of our results are very much appreciated.
This work was partially supported by the Russian Science Foundation (grant 21-72-20067).
We acknowledge the discussions with participants
of the Nordita Scientific Program on ''Towards a comprehensive model
of the galactic magnetic field", Stockholm (April 2023),
which is partly supported by NordForsk.

\section*{Data Availability}

\noindent
There are no new data associated with this article.

\bibliographystyle{mnras}
\bibliography{Budget-Eqns-MNRAS}

\begin{thebibliography}{}
\makeatletter
\relax
\def\mn@urlcharsother{\let\do\@makeother \do\$\do\&\do\#\do\^\do\_\do\%\do\~}
\def\mn@doi{\begingroup\mn@urlcharsother \@ifnextchar [ {\mn@doi@}
  {\mn@doi@[]}}
\def\mn@doi@[#1]#2{\def\@tempa{#1}\ifx\@tempa\@empty \href
  {http://dx.doi.org/#2} {doi:#2}\else \href {http://dx.doi.org/#2} {#1}\fi
  \endgroup}
\def\mn@eprint#1#2{\mn@eprint@#1:#2::\@nil}
\def\mn@eprint@arXiv#1{\href {http://arxiv.org/abs/#1} {{\tt arXiv:#1}}}
\def\mn@eprint@dblp#1{\href {http://dblp.uni-trier.de/rec/bibtex/#1.xml}
  {dblp:#1}}
\def\mn@eprint@#1:#2:#3:#4\@nil{\def\@tempa {#1}\def\@tempb {#2}\def\@tempc
  {#3}\ifx \@tempc \@empty \let \@tempc \@tempb \let \@tempb \@tempa \fi \ifx
  \@tempb \@empty \def\@tempb {arXiv}\fi \@ifundefined
  {mn@eprint@\@tempb}{\@tempb:\@tempc}{\expandafter \expandafter \csname
  mn@eprint@\@tempb\endcsname \expandafter{\@tempc}}}

\bibitem[\protect\citeauthoryear{Baryshnikova \& Shukurov}{Baryshnikova \&
  Shukurov}{1987}]{BS87}
Baryshnikova Y.,  Shukurov A.,  1987, Astron. Nachr., 308, 89

\bibitem[\protect\citeauthoryear{Bassom, Kuzanyan, Sokoloff  \& Soward}{Bassom
  et~al.}{2005}]{BSK05}
Bassom A.~P.,  Kuzanyan K.~M.,  Sokoloff D.,   Soward A.~M.,  2005, Geophysical
  and Astrophysical Fluid Dynamics, 99, 309

\bibitem[\protect\citeauthoryear{Blackman \& Brandenburg}{Blackman \&
  Brandenburg}{2002}]{BB02}
Blackman E.~G.,  Brandenburg A.,  2002, Astrophys. J., 579, 359

\bibitem[\protect\citeauthoryear{Blackman \& Field}{Blackman \&
  Field}{2000}]{BF00}
Blackman E.~G.,  Field G.~B.,  2000, Astrophys. J., 534, 984

\bibitem[\protect\citeauthoryear{Brandenburg}{Brandenburg}{2017}]{B17}
Brandenburg A.,  2017, Astron. Astrophys., 598, A117

\bibitem[\protect\citeauthoryear{Brandenburg \& Subramanian}{Brandenburg \&
  Subramanian}{2005}]{Brandenburg(2005)}
Brandenburg A.,  Subramanian K.,  2005, Phys. Rep., 417, 1

\bibitem[\protect\citeauthoryear{Chamandy \& Singh}{Chamandy \&
  Singh}{2018}]{CS18}
Chamandy L.,  Singh N.~K.,  2018, MNRAS, 481, 1300

\bibitem[\protect\citeauthoryear{Chamandy, Shukurov, Subramanian  \&
  Stoker}{Chamandy et~al.}{2014}]{CSS14}
Chamandy L.,  Shukurov A.,  Subramanian K.,   Stoker K.,  2014, MNRAS, 443,
  1867

\bibitem[\protect\citeauthoryear{Covas, Tworkowski, Tavakol  \&
  Brandenburg}{Covas et~al.}{1997}]{CTB97}
Covas E.,  Tworkowski A.,  Tavakol R.,   Brandenburg A.,  1997, Solar Phys.,
  172, 3

\bibitem[\protect\citeauthoryear{Covas, Tavakol, Tworkowski  \&
  Brandenburg}{Covas et~al.}{1998}]{CTB98}
Covas E.,  Tavakol R.,  Tworkowski A.,   Brandenburg A.,  1998, Astron.
  Astrophys., 329, 350

\bibitem[\protect\citeauthoryear{Del~Sordo, Guerrero  \& Brandenburg}{Del~Sordo
  et~al.}{2013}]{DGB13}
Del~Sordo F.,  Guerrero G.,   Brandenburg A.,  2013, Monthly Notices of the
  Royal Astronomical Society, 429, 1686

\bibitem[\protect\citeauthoryear{Elperin, Kleeorin, Rogachevskii  \&
  Zilitinkevich}{Elperin et~al.}{2002}]{EKRZ02}
Elperin T.,  Kleeorin N.,  Rogachevskii I.,   Zilitinkevich S.,  2002, Phys.
  Rev. E, 66, 066305

\bibitem[\protect\citeauthoryear{Field, Blackman  \& Chou}{Field
  et~al.}{1999}]{FBC99}
Field G.~B.,  Blackman E.~G.,   Chou H.,  1999, Astrophys. J., 513, 638

\bibitem[\protect\citeauthoryear{Gopalakrishnan \& Subramanian}{Gopalakrishnan
  \& Subramanian}{2023}]{GS23}
Gopalakrishnan K.,  Subramanian K.,  2023, Astrophys. J., 8, 65

\bibitem[\protect\citeauthoryear{Griffiths, Bassom, Soward  \&
  Kuzanyan}{Griffiths et~al.}{2001}]{GSK01}
Griffiths G.~L.,  Bassom A.~P.,  Soward A.~M.,   Kuzanyan K.~M.,  2001,
  Geophysical and Astrophysical Fluid Dynamics, 94, 85

\bibitem[\protect\citeauthoryear{Gruzinov \& Diamond}{Gruzinov \&
  Diamond}{1994}]{GD94}
Gruzinov A.~V.,  Diamond P.~H.,  1994, Phys. Rev. Lett., 72, 1651

\bibitem[\protect\citeauthoryear{Guerrero, Chatterjee  \& Brandenburg}{Guerrero
  et~al.}{2010}]{GCB10}
Guerrero G.,  Chatterjee P.,   Brandenburg A.,  2010, MNRAS, 409, 1619

\bibitem[\protect\citeauthoryear{Hubbard \& Brandenburg}{Hubbard \&
  Brandenburg}{2012}]{HB12}
Hubbard A.,  Brandenburg A.,  2012, Astrophys. J., 748, 51

\bibitem[\protect\citeauthoryear{K{\"a}pyl{\"a}, Korpi  \&
  Brandenburg}{K{\"a}pyl{\"a} et~al.}{2010}]{KKB10}
K{\"a}pyl{\"a} P.~J.,  Korpi M.~J.,   Brandenburg A.,  2010, Astron.
  Astrophys., 518, A22

\bibitem[\protect\citeauthoryear{Kitchatinov, Pipin  \&
  R{\"u}diger}{Kitchatinov et~al.}{1994}]{KRP94}
Kitchatinov L.~L.,  Pipin V.~V.,   R{\"u}diger G.,  1994, Astron. Nachr., 315,
  157

\bibitem[\protect\citeauthoryear{Kleeorin \& Rogachevskii}{Kleeorin \&
  Rogachevskii}{1999}]{KR99}
Kleeorin N.,  Rogachevskii I.,  1999, Phys. Rev. E, 59, 6724

\bibitem[\protect\citeauthoryear{Kleeorin \& Rogachevskii}{Kleeorin \&
  Rogachevskii}{2022}]{KR22}
Kleeorin N.,  Rogachevskii I.,  2022, Monthly Notices of the Royal Astronomical
  Society, 515, 5437

\bibitem[\protect\citeauthoryear{Kleeorin \& Ruzmaikin}{Kleeorin \&
  Ruzmaikin}{1982}]{KR82}
Kleeorin N.,  Ruzmaikin A.,  1982, Magnetohydrodynamics, 18, 116

\bibitem[\protect\citeauthoryear{Kleeorin, Rogachevskii  \& Ruzmaikin}{Kleeorin
  et~al.}{1990}]{KRR90}
Kleeorin N.,  Rogachevskii I.,   Ruzmaikin A.~A.,  1990, Sov. Phys. JETP, 70,
  878

\bibitem[\protect\citeauthoryear{Kleeorin, Rogachevskii  \& Ruzmaikin}{Kleeorin
  et~al.}{1995}]{KRR95}
Kleeorin N.,  Rogachevskii I.,   Ruzmaikin A.,  1995, Astron. Astrophys., 297,
  159

\bibitem[\protect\citeauthoryear{Kleeorin, Moss, Rogachevskii  \&
  Sokoloff}{Kleeorin et~al.}{2000}]{KMR00}
Kleeorin N.,  Moss D.,  Rogachevskii I.,   Sokoloff D.,  2000, Astron.
  Astrophys., 361, L5

\bibitem[\protect\citeauthoryear{Kleeorin, Moss, Rogachevskii  \&
  Sokoloff}{Kleeorin et~al.}{2002}]{KMR02}
Kleeorin N.,  Moss D.,  Rogachevskii I.,   Sokoloff D.,  2002, Astron.
  Astrophys., 387, 453

\bibitem[\protect\citeauthoryear{Kleeorin, Moss, Rogachevskii  \&
  Sokoloff}{Kleeorin et~al.}{2003a}]{KMR03a}
Kleeorin N.,  Moss D.,  Rogachevskii I.,   Sokoloff D.,  2003a, Astron.
  Astrophys., 400, 9

\bibitem[\protect\citeauthoryear{Kleeorin, Kuzanyan, Moss, Rogachevskii,
  Sokoloff  \& Zhang}{Kleeorin et~al.}{2003b}]{KKMR03}
Kleeorin N.,  Kuzanyan K.,  Moss D.,  Rogachevskii I.,  Sokoloff D.,   Zhang
  H.,  2003b, Astron. Astrophys., 409, 1097

\bibitem[\protect\citeauthoryear{Kleeorin, Safiullin, Kleeorin, Porshnev,
  Rogachevskii  \& Sokoloff}{Kleeorin et~al.}{2016}]{KSR16}
Kleeorin Y.,  Safiullin N.,  Kleeorin N.,  Porshnev S.,  Rogachevskii I.,
  Sokoloff D.,  2016, Mon. Not. Roy. Astron. Soc., 460, 3960

\bibitem[\protect\citeauthoryear{Kleeorin, Safiullin, Kuzanyan, Rogachevskii,
  Tlatov  \& Porshnev}{Kleeorin et~al.}{2020}]{KSR20}
Kleeorin N.,  Safiullin N.,  Kuzanyan K.,  Rogachevskii I.,  Tlatov A.,
  Porshnev S.,  2020, Mon. Not. Roy. Astron. Soc., 495, 238

\bibitem[\protect\citeauthoryear{Kleeorin, Rogachevskii, Safiullin, Gershberg
  \& Porshnev}{Kleeorin et~al.}{2023}]{KRS23}
Kleeorin N.,  Rogachevskii I.,  Safiullin N.,  Gershberg R.,   Porshnev S.,
  2023, MNRAS, 526, 1601

\bibitem[\protect\citeauthoryear{Krause \& R{\"a}dler}{Krause \&
  R{\"a}dler}{1980}]{Krause(1980)}
Krause F.,  R{\"a}dler K.-H.,  1980, Mean-Field Magnetohydrodynamics and Dynamo
  Theory.
Oxford: Pergamon Press

\bibitem[\protect\citeauthoryear{McComb}{McComb}{1990}]{MC90}
McComb W.~D.,  1990, The Physics of Fluid Turbulence.
Oxford Science Publications

\bibitem[\protect\citeauthoryear{Meunier, Nesme-Ribes  \& Sokoloff}{Meunier
  et~al.}{1996}]{MNS96}
Meunier N.,  Nesme-Ribes E.,   Sokoloff D.,  1996, Astron. Rep., 40, 415

\bibitem[\protect\citeauthoryear{Moffatt}{Moffatt}{1978}]{Moffatt(1978)}
Moffatt H.~K.,  1978, Magnetic Field Generation in Electrically Conducting
  Fluids.
Cambridge: Cambridge University Press

\bibitem[\protect\citeauthoryear{Moffatt \& Dormy}{Moffatt \&
  Dormy}{2019}]{MD2019}
Moffatt H.~K.,  Dormy E.,  2019, Self-Exciting Fluid Dynamos.
Cambridge: Cambridge University Press

\bibitem[\protect\citeauthoryear{Monin \& Yaglom}{Monin \& Yaglom}{1971}]{MY71}
Monin A.~S.,  Yaglom A.~M.,  1971, Statistical Fluid Mechanics, Vol. 1.
Cambridge, Mass: MIT Press

\bibitem[\protect\citeauthoryear{Monin \& Yaglom}{Monin \& Yaglom}{2013}]{MY13}
Monin A.~S.,  Yaglom A.~M.,  2013, Statistical Fluid Mechanics, Vol. 2.
New York: Dover

\bibitem[\protect\citeauthoryear{Orszag}{Orszag}{1970}]{O70}
Orszag S.~A.,  1970, J. Fluid Mech., 41, 363

\bibitem[\protect\citeauthoryear{Parker}{Parker}{1979}]{Parker(1979)}
Parker E.~N.,  1979, Cosmical Magnetic Fields: Their Origin and their Activity.
Oxford: Clarendon Press

\bibitem[\protect\citeauthoryear{Pouquet, Frisch  \& L{\'e}orat}{Pouquet
  et~al.}{1976}]{PFL76}
Pouquet A.,  Frisch U.,   L{\'e}orat J.,  1976, J. Fluid Mech., 77, 321

\bibitem[\protect\citeauthoryear{R{\"a}dler \& Br{\"a}uer}{R{\"a}dler \&
  Br{\"a}uer}{1987}]{RB87}
R{\"a}dler K.-H.,  Br{\"a}uer H.-J.,  1987, Astron. Nachr., 308, 101

\bibitem[\protect\citeauthoryear{Rincon}{Rincon}{2021}]{RIN21}
Rincon F.,  2021, Phys. Rev. Fluids, 6, L121701

\bibitem[\protect\citeauthoryear{Roberts \& Soward}{Roberts \&
  Soward}{1975}]{RS75}
Roberts P.~H.,  Soward A.~M.,  1975, Astron. Nachr., 296, 49

\bibitem[\protect\citeauthoryear{Rogachevskii}{Rogachevskii}{2021}]{RI21}
Rogachevskii I.,  2021, Introduction to Turbulent Transport of Particles,
  Temperature and Magnetic Fields.
Cambridge: Cambridge University Press

\bibitem[\protect\citeauthoryear{Rogachevskii \& Kleeorin}{Rogachevskii \&
  Kleeorin}{2000}]{RK2000}
Rogachevskii I.,  Kleeorin N.,  2000, Phys. Rev. E, 61, 5202

\bibitem[\protect\citeauthoryear{Rogachevskii \& Kleeorin}{Rogachevskii \&
  Kleeorin}{2001}]{RK2001}
Rogachevskii I.,  Kleeorin N.,  2001, Phys. Rev. E, 64, 056307

\bibitem[\protect\citeauthoryear{Rogachevskii \& Kleeorin}{Rogachevskii \&
  Kleeorin}{2004}]{RK2004}
Rogachevskii I.,  Kleeorin N.,  2004, Phys. Rev. E, 70, 046310

\bibitem[\protect\citeauthoryear{Rogachevskii \& Kleeorin}{Rogachevskii \&
  Kleeorin}{2006}]{RK2006}
Rogachevskii I.,  Kleeorin N.,  2006, Geophys. Astrophys. Fluid Dyn., 100, 243

\bibitem[\protect\citeauthoryear{Rogachevskii \& Kleeorin}{Rogachevskii \&
  Kleeorin}{2007}]{RK2007}
Rogachevskii I.,  Kleeorin N.,  2007, Phys. Rev. E, 76, 056307

\bibitem[\protect\citeauthoryear{R{\"u}diger \& Kichatinov}{R{\"u}diger \&
  Kichatinov}{1993}]{RK93}
R{\"u}diger G.,  Kichatinov L.~L.,  1993, Astron. Astrophys., 269, 581

\bibitem[\protect\citeauthoryear{R{\"u}diger, Hollerbach  \&
  Kitchatinov}{R{\"u}diger et~al.}{2013}]{Ruediger(2013)}
R{\"u}diger G.,  Hollerbach R.,   Kitchatinov L.~L.,  2013, Magnetic Processes
  in Astrophysics: Theory, Simulations, Experiments.
Weinheim: John Wiley \& Sons

\bibitem[\protect\citeauthoryear{Ruzmaikin, Shukurov  \& Sokoloff}{Ruzmaikin
  et~al.}{1988}]{Ruzmaikin(1988)}
Ruzmaikin A.,  Shukurov A.~M.,   Sokoloff D.~D.,  1988, Magnetic Fields of
  Galaxies.
Dordrecht: Kluwer Academic

\bibitem[\protect\citeauthoryear{Safiullin, Kleeorin, Porshnev, Rogachevskii
  \& Ruzmaikin}{Safiullin et~al.}{2018}]{SKR18}
Safiullin N.,  Kleeorin N.,  Porshnev S.,  Rogachevskii I.,   Ruzmaikin A.,
  2018, J. Plasma Phys., 84, 735840306

\bibitem[\protect\citeauthoryear{Shukurov \& Subramanian}{Shukurov \&
  Subramanian}{2021}]{SS21}
Shukurov A.,  Subramanian K.,  2021, Astrophysical Magnetic Fields: From
  Galaxies to the Early Universe.
Cambridge University Press

\bibitem[\protect\citeauthoryear{Shukurov, Sokoloff  \& Ruzmaikin}{Shukurov
  et~al.}{1985}]{SSR85}
Shukurov A.~M.,  Sokoloff D.~D.,   Ruzmaikin A.~A.,  1985, Magnitnaya
  Gidrodynamika, 1, 9

\bibitem[\protect\citeauthoryear{Shukurov, Sokoloff, Subramanian  \&
  Brandenburg}{Shukurov et~al.}{2006}]{SSS06}
Shukurov A.,  Sokoloff D.,  Subramanian K.,   Brandenburg A.,  2006, Astron.
  Astrophys., 448, L33

\bibitem[\protect\citeauthoryear{Sokoloff, Bao, Kleeorin, Kuzanyan, Moss,
  Rogachevskii, Tomin  \& Zhang}{Sokoloff et~al.}{2006}]{SKR06}
Sokoloff D.,  Bao S.~D.,  Kleeorin N.,  Kuzanyan K.,  Moss D.,  Rogachevskii
  I.,  Tomin D.,   Zhang H.,  2006, Astron. Nachr., 327, 876

\bibitem[\protect\citeauthoryear{Subramanian \& Brandenburg}{Subramanian \&
  Brandenburg}{2004}]{SB04}
Subramanian K.,  Brandenburg A.,  2004, Phys. Rev. Lett., 93, 205001

\bibitem[\protect\citeauthoryear{Vishniac \& Cho}{Vishniac \& Cho}{2001}]{VC01}
Vishniac E.~T.,  Cho J.,  2001, Astrophys. J., 550, 752

\bibitem[\protect\citeauthoryear{Zeldovich, Ruzmaikin  \& Sokoloff}{Zeldovich
  et~al.}{1983}]{Zeldovich(1983)}
Zeldovich Y.~B.,  Ruzmaikin A.~A.,   Sokoloff D.~D.,  1983, Magnetic Fields in
  Astrophysics.
New-York: Gordon and Breach

\bibitem[\protect\citeauthoryear{Zhang, Sokoloff, Rogachevskii, Moss, Lamburt,
  Kuzanyan  \& Kleeorin}{Zhang et~al.}{2006}]{ZKRS06}
Zhang H.,  Sokoloff D.,  Rogachevskii I.,  Moss D.,  Lamburt V.,  Kuzanyan K.,
   Kleeorin N.,  2006, Mon. Not. Roy. Astron. Soc., 365, 276

\bibitem[\protect\citeauthoryear{Zhang, Moss, Kleeorin, Kuzanyan, Rogachevskii,
  Sokoloff, Gao  \& Xu}{Zhang et~al.}{2012}]{ZKRS12}
Zhang H.,  Moss D.,  Kleeorin N.,  Kuzanyan K.,  Rogachevskii I.,  Sokoloff D.,
   Gao Y.,   Xu H.,  2012, Astrophys. J., 751, 47

\makeatother
\end{thebibliography}

\appendix

\section{Quenching functions}
\label{sect-A1}

The quenching functions $\phi_{_{\rm K}}(\beta)$ and $\phi_{_{\rm M}}(\beta)$ are
\begin{eqnarray}
\phi_{_{\rm K}}(\beta) &=& A_{1}^{(1)}\left(\sqrt{2} \beta\right)
+ A_{2}^{(1)}\left(\sqrt{2} \beta\right) = {1 \over 7} \, \biggl\{
3 \biggl[1 - 4 \beta^{2}
\nonumber\\
&&+ 8 \beta^{4} \ln \left(1 + (2\beta^2)^{-1}\right)\biggr] + 4 \phi_{_{\rm M}}\left(\sqrt{2} \,\beta\right)\biggr\},
\label{WECB41}
\end{eqnarray}
\begin{eqnarray}
\phi_{_{\rm M}}(\beta) &=& {3 \over 4 \pi} \, \left[\tilde A_{1}\left(\beta^2\right)
+ \tilde A_{2}\left(\beta^2\right)\right]
\nonumber\\
&=& {3 \over \beta^{2}} \, \left(1 - {\arctan \beta \over \beta}\right) .
\label{WECB40}
\end{eqnarray}
The quenching function $\phi_\eta^{(B)}(\beta)$ is given by
$\phi_\eta^{(B)}(\beta)=\phi_{_{\rm K}}(\beta) + \phi(\beta)$ and
\begin{eqnarray}
\phi(\beta) &=& (2 - 3\epsilon) \, A_{2}^{(1)}\left(\sqrt{2} \beta\right)
- {3 \over 2 \pi} \, (1 - \epsilon) \, \tilde A_{2}\left(2 \beta^2\right) ,
\label{WECB402}
\end{eqnarray}
where the functions $A_{1}^{(1)}(\beta) $ and $A_{2}^{(1)}(\beta) $ are
given by
\begin{eqnarray}
A_{1}^{(1)}(\beta) &=& {6 \over 5} \biggl[{\arctan \beta \over
\beta} \biggl(1 + {5 \over 7 \beta^{2}} \biggr) + {1 \over 14}
L(\beta) - {5 \over 7\beta^{2}} \biggr] ,
\nonumber \\
\label{X61}\\
A_{2}^{(1)}(\beta) &=& - {6 \over 5} \biggl[{\arctan \beta \over
\beta} \biggl(1 + {15 \over 7 \beta^{2}} \biggr) - {2 \over 7}
L(\beta) - {15 \over 7\beta^{2}} \biggr]  ,
\nonumber \\
\label{X62}
\end{eqnarray}
and $ L(\beta) = 1 - 2 \beta^{2} + 2 \beta^{4} \ln (1 + \beta^{-2})$.
For $ \beta \ll 1 $ these functions are given by
\begin{eqnarray*}
A_{1}^{(1)}(\beta) &\sim& 1 - {2 \over 5} \beta^{2}  \;, \quad
A_{2}^{(1)}(\beta) \sim - {4 \over 5} \beta^{2} ,
\end{eqnarray*}
and for $ \beta \gg 1 $ they are given by
\begin{eqnarray*}
A_{1}^{(1)}(\beta) &\sim& {3 \pi \over 5 \beta} - {2 \over
\beta^{2}} \;, \quad A_{2}^{(1)}(\beta) \sim - {3 \pi \over 5
\beta} + {4 \over \beta^{2}} .
\end{eqnarray*}
The functions $\tilde A_{1}(x)$ and $\tilde A_{2}(x)$ are
given by
\begin{eqnarray}
\tilde A_{1}(x) &=& {2 \pi \over x} \biggl[(x + 1) {\arctan (\sqrt{x}) \over
\sqrt{x}} - 1 \biggr] ,
\label{XXX61}\\
\tilde A_{2}(x) &=& - {2 \pi \over x} \biggl[(x + 3) {\arctan (\sqrt{x}) \over
\sqrt{x}} - 3 \biggr] .
\label{XXX62}
\end{eqnarray}
For $x \ll 1 $ these functions are given by
\begin{eqnarray*}
\tilde A_{1}(x) &\sim& {4 \pi \over 3} \biggl(1 - {1 \over 5} x
\biggr) , \quad \tilde A_{2}(x) \sim - {8 \pi \over 15} x ,
\end{eqnarray*}
In the case of $x \gg 1 $ these functions are given by
\begin{eqnarray*}
\tilde A_{1}(x) &\sim& {\pi^{2} \over \sqrt{x}} - {4 \pi \over x}
\;, \quad \tilde A_{2}(x) \sim - {\pi^{2} \over \sqrt{x}} + {8 \pi
\over x} .
\end{eqnarray*}

\section{Derivation of Eqs.~(\ref{C3})--(\ref{C5})}
\label{sect-A2}

In this Appendix we derive Eqs.~(\ref{C3})--(\ref{C5})
[for more details see paper by~\cite{RK2007}].
Using procedure described in Section~\ref{sect-2}, we derive
equations for the correlation functions of the velocity fluctuations
$f_{ij} = \langle u_i \, u_j\rangle$, the magnetic fluctuations
$h_{ij} = \langle b_i \, b_j \rangle$,  and the cross-helicity $g_{ij} =
\langle u_i \, b_j \rangle$ in the Fourier space:
\begin{eqnarray}
{\partial f_{ij}({\bm k}) \over \partial t} = -{\rm i}\,({\bm k}
{\bm \cdot} \meanBB) \, \Phi_{ij}({\bm k}) + \hat{\cal M} f^{(III)}_{ij}({\bm k}) ,
\label{WB6}
\end{eqnarray}
\begin{eqnarray}
{\partial h_{ij}({\bm k}) \over \partial t} = {\rm i}\,({\bm
k}{\bm \cdot} \meanBB) \Phi_{ij}({\bm k}) + \hat{\cal M} h^{(III)}_{ij}({\bm k}) ,
\label{WB7}
\end{eqnarray}
\begin{eqnarray}
{\partial g_{ij}({\bm k}) \over \partial t} = - {\rm i}\,\left({\bm k}
{\bm \cdot} \meanBB\right) \, \left[f_{ij}({\bm k}) - h_{ij}({\bm k})\right]
+ \hat{\cal M} g^{(III)}_{ij}({\bm k}) ,
\label{WB8}
\end{eqnarray}
where $\Phi_{ij}({\bm k}) = g_{ij}({\bm k}) - g_{ji}(-{\bm k})$,
and $ \hat{\cal M} f^{(III)}_{ij}$, $\, \hat{\cal M}h^{(III)}_{ij}$,
and $\hat{\cal M}g^{(III)}_{ij}$ are the third-order
moment terms appearing due to the nonlinear terms.
We split the tensor $\langle b_i \, b_j \rangle$
of magnetic fluctuations into
nonhelical $h_{ij}$ and helical $h_{ij}^{(H)}$ parts.
The helical part $h_{ij}^{(H)}$ depends on the magnetic helicity, and it is
determined by the dynamic equation which follows from the magnetic
helicity conservation arguments.
We also split the second-order
correlation functions into symmetric and antisymmetric parts with
respect to the wave vector ${\bm k}$, e.g., $f_{ij} = f_{ij}^{(s)} +
f_{ij}^{(a)}$, where the tensors $f_{ij}^{(s)} = [f_{ij}({\bm k}) +
f_{ij}(-{\bm k})] / 2$ describes the symmetric part of the tensor
and $f_{ij}^{(a)} = [f_{ij}({\bm k}) - f_{ij}(-{\bm k})] / 2$
determines the antisymmetric part of the tensor.
We apply the spectral $\tau$ approximation [see Eq.~(\ref{WEA1})]
for the nonhelical parts of the tensors.
We assume that the characteristic time of variation of the mean
magnetic field $\meanBB$ is substantially larger than the
correlation time $\tilde \tau(k)$ for all turbulence scales. This allows us
to get a stationary solution for the equations for the second-order
moments
\begin{eqnarray}
f_{ij}^{(s)}({\bm k}) = {1 \over 1 + 2 \psi} \, \left[(1 + \psi)
\, f_{ij}^{(0s)}({\bm k}) + \psi \, h_{ij}^{(0s)}({\bm k})\right]  ,
\label{ZB22}
\end{eqnarray}
\begin{eqnarray}
h_{ij}^{(s)}({\bm k}) = {1 \over 1 + 2 \psi} \,  \left[\psi
\, f_{ij}^{(0s)}({\bm k}) + (1 + \psi) \, h_{ij}^{(0s)}({\bm k})\right] ,
\label{ZB24}
\end{eqnarray}
where $ \psi({\bm k}) = 2 (\tilde\tau \,{\bm k} {\bm \cdot} \meanBB)^2$.
Next, we specify a model for the background turbulence (with
zero mean magnetic field $\meanBB = 0)$ [denoted with the
superscript $(0)$], see Eqs.~(\ref{WEM13})--(\ref{WEMM13}).
The background turbulence here is assumed to be homogeneous, isotropic and non-helical.
Integration in ${\bm k}$ space in Eqs.~(\ref{ZB22})--(\ref{ZB24}) yields
Eqs.~(\ref{C3})--(\ref{C5}), where
the nonlinear functions $q_{\rm p}(\beta) $ and $q_{\rm s}(\beta)$ are given by
\begin{eqnarray}
q_{\rm p}(\beta) &=& {2\over 3\beta^2} \, \left[A_{1}^{(0)}(0) - A_{1}^{(0)}(\sqrt{2}\beta)
- A_{2}^{(0)}(\sqrt{2}\beta)\right] ,
\label{X1}\\
q_{\rm s}(\beta) &=& - {2\over 3\beta^2} \,  A_{2}^{(0)}(\sqrt{2}\beta) ,
\label{X2}
\end{eqnarray}
and $\beta = \sqrt{8} \,\,  \meanB/ \meanB_{\rm eq}$.
The functions $A_{1}^{(0)}(\beta)$ and $A_{2}^{(0)}(\beta)$ are given by
\begin{eqnarray}
A_{1}^{(0)}(\beta) &=& {1 \over 5} \biggl[2 + 2 {\arctan \beta
\over \beta^3} (3 + 5 \beta^{2}) - {6 \over \beta^{2}}  -
\beta^{2} \ln {\rm Rm}
\nonumber \\
& & - 2 \beta^{2} \ln \biggl({1 + \beta^{2} \over 1 + \beta^{2}
\sqrt{\rm Rm}}\biggr) \biggr] \;,
\label{X40}\\
A_{2}^{(0)}(\beta) &=& {2 \over 5} \biggl[2 - {\arctan \beta \over
\beta^3} (9 + 5 \beta^{2}) + {9 \over \beta^{2}}  - \beta^{2} \ln
{\rm Rm}
\nonumber \\
& & - 2 \beta^{2} \ln \biggl({1 + \beta^{2} \over 1 + \beta^{2}
\sqrt{\rm Rm}}\biggr) \biggr] .
\label{X41}
\end{eqnarray}
For $\meanB \ll \meanB_{\rm eq} / 4 {\rm Rm}^{1/4} $, these functions are given by
\begin{eqnarray*}
A_{1}^{(0)}(\beta) &\sim& 2 - {1 \over 5} \beta^{2} \ln {\rm Rm} ,
\\
A_{2}^{(0)}(\beta) &\sim& - {2 \over 5} \beta^{2} \biggl[\ln {\rm
Rm} +  {2 \over 15}\biggr]\; .
\end{eqnarray*}
For $\meanB_{\rm eq} / 4 {\rm Rm}^{1/4} \ll  \meanB \ll \meanB_{\rm eq} / 4$, these functions are given by
\begin{eqnarray*}
A_{1}^{(0)}(\beta) &\sim& 2 + {2 \over 5} \beta^{2} \biggl[2 \ln
\beta -  {16 \over 15} + {4 \over 7} \beta^{2} \biggr] \;,
\\
A_{2}^{(0)}(\beta) &\sim& {2 \over 5} \beta^{2} \biggl[4 \ln \beta
-  {2 \over 15} - 3 \beta^{2} \biggr] \;,
\end{eqnarray*}
and for $\meanB \gg \meanB_{\rm eq} / 4 $, they are given by
\begin{eqnarray*}
A_{1}^{(0)}(\beta) &\sim& {\pi \over \beta} - {3 \over \beta^{2}}
\;, \quad A_{2}^{(0)}(\beta) \sim - {\pi \over \beta} + {6 \over
\beta^{2}} .
\end{eqnarray*}

\end{document}